\documentclass[longauth]{aa}
\usepackage[switch]{lineno}
\usepackage{graphicx}
\usepackage{subcaption}
\usepackage{placeins} 
\usepackage[dvipsnames]{xcolor}
\usepackage{txfonts}
\usepackage[flushleft]{threeparttable}
\usepackage{float}
\usepackage{graphicx}
\graphicspath{{graphics}}
\usepackage{array}
\usepackage{arydshln}
\usepackage{multirow}
\usepackage[colorlinks=true,linkcolor=blue,citecolor=blue]{hyperref}
\def\Msol{{\rm M}_\odot}
\def\Mstar{{\rm M}_\star}

\def\Phidt{\Phi_{\Delta t }}

\def\phiseven{\Phi_{700}}
\def\phigiga{\Phi_{1000}}
\def\T50{T$_{50\%}$}

\def\rdt{\textrm{r}_{\Delta t }}

\def\rfive{\textrm{r}_{500}}
\def\rseven{\textrm{r}_{700}}
\def\rgiga{\textrm{r}_{1000}}

\def\SFRM{\textrm{M}_\star-\textrm{SFR}}

\def \cigale{{\texttt{CIGALE}}}
\def\photoz{\mbox{photo-\textit{z}}}
\def\specz{\mbox{spec-\textit{z}}}

\def\lephare{\texttt{LePHARE}}
\def\sigmaz{\sigma_{\Delta z / (1+z_{\rm s})}}
\def\migvec{\overrightarrow{m}_{\Delta t}}

\definecolor{Blue}{rgb}{0,0.25,0.9}
\definecolor{Red}{rgb}{0.85,0.08,0.05}
\definecolor{Green}{rgb}{0.35,0.45,0.25}
\definecolor{Orange}{rgb}{1.0,0.5,0.15}
\definecolor{Brown}{rgb}{0.7,0.25,0.0}

\begin{document}

   \title{COSMOS-Web: A history of galaxy migrations over the stellar mass–star formation rate plane
}

   \subtitle{}

   \author{R. C. Arango-Toro\inst{1}\fnmsep\thanks{E-mail: \url{rafael.arango-toro@lam.fr}}, 
    O.~Ilbert\inst{\ref{LAM}},
    L.~Ciesla\inst{\ref{LAM}},
    M.~Shuntov \inst{\ref{DAWN},\ref{UOC}},
    G.~Aufort\inst{\ref{IAP}},
    W.~Mercier\inst{\ref{LAM}}
    C.~Laigle \inst{\ref{IAP}},
    M.~Franco \inst{\ref{UAT}},
    M.~Bethermin\inst{\ref{OAS}},
    D.~Le Borgne\inst{\ref{IAP}},
    Y.~Dubois \inst{\ref{IAP}},
    H. J. McCracken \inst{\ref{IAP}},
    L. Paquereau\inst{\ref{IAP}},
    M.~Huertas-Company\inst{\ref{IAC},\ref{Lerma},\ref{Uparis}},
    J.~Kartaltepe\inst{\ref{Rochester}},
    C. M. ~Casey\inst{\ref{UAT},\ref{DAWN}},
    H.~Akins\inst{\ref{UAT}}
    N.~Allen\inst{\ref{DAWN}}
    I.~Andika\inst{\ref{TUM},\ref{MPI}},
    M.~Brinch\inst{\ref{Valpa}},
    N. E.~Drakos\inst{\ref{HawaiiHilo}},
    A.~Faisst\inst{\ref{Caltech}},
    G.~Gozaliasl\inst{\ref{DCS},\ref{Helsinki}},
    S.~Harish\inst{\ref{Rochester}},
    A.~Kaminsky\inst{\ref{SOFIA}},
    A.~Koekemoer\inst{\ref{STScI}},
    V.~Kokorev\inst{\ref{Groningen}},
    D.~Liu\inst{\ref{Nanjing}},
    G.~Magdis\inst{\ref{DAWN},\ref{UOC},\ref{DTU}},
    C. L. ~Martin\inst{\ref{SantaBarbara}},
    T.~Moutard\inst{\ref{ESAC}},
    J.~Rhodes\inst{\ref{NASA}},
    R. M.~Rich\inst{\ref{Losangeles}},
    B.~Robertson\inst{\ref{UnivCalifornia}},
    D. B.~Sanders\inst{\ref{HawaiiHonolulu}},
    K.~Sheth\inst{\ref{Jack-NASA}},
    M.~Talia\inst{\ref{Bologna},\ref{INAF}},
    S.~Toft\inst{\ref{DAWN},\ref{UOC}},
    L.~Tresse\inst{\ref{LAM}},
    F.~Valentino\inst{\ref{DAWN},\ref{ESO-Ger}},
    A.~Vijayan\inst{\ref{Brighton}},
    J.~Weaver\inst{\ref{UMASS}}
    }

\institute{
    Aix Marseille Univ, CNRS, CNES, LAM, Marseille, France\label{LAM}%
    \and%
    Cosmic Dawn Center (DAWN), Denmark\label{DAWN}%
    \and%
    Niels Bohr Institute, University of Copenhagen, Jagtvej 128, DK-2200, Copenhagen, Denmark\label{UOC}%
    \and%
    Institut d’Astrophysique de Paris, UMR 7095, CNRS, Sorbonne Université, 98 bis boulevard Arago, F-75014 Paris, France\label{IAP}%
    \and%
    The University of Texas at Austin, 2515 Speedway Blvd Stop C1400, Austin, TX 78712, USA\label{UAT}%
    \and%
    Université de Strasbourg, CNRS, Observatoire astronomique de Strasbourg, UMR 7550, 67000 Strasbourg, France\label{OAS}%
    \and%
    Instituto de Astrofísica de Canarias (IAC), La Laguna, E-38205, Spain\label{IAC}%
    \and%
    Observatoire de Paris, LERMA, PSL University, 61 avenue de l’Observatoire, F-75014 Paris, France\label{Lerma}%
    \and%
    Université Paris-Cité, 5 Rue Thomas Mann, 75014 Paris, France\label{Uparis}%
    \and%
    Laboratory for Multiwavelength Astrophysics, School of Physics and Astronomy, Rochester Institute of Technology, 84 Lomb Memorial Drive, Rochester, NY 14623, USA\label{Rochester}%
    \and%
    Technical University of Munich, TUM School of Natural Sciences, Department of Physics, James-Franck-Str. 1, D-85748 Garching, Germany\label{TUM}%
    \and%
    Max-Planck-Institut f\"{u}r Astrophysik, Karl-Schwarzschild-Str. 1, D-85748 Garching, Germany\label{MPI}%
    \and%
    Instituto de Física y Astronomía, Universidad de Valparaíso, Avda. Gran Bretana~ 1111, Valparaíso, Chile \label{Valpa}
    \and%
    Department of Physics and Astronomy, University of Hawaii, Hilo, 200 W Kawili St, Hilo, HI 96720, USA\label{HawaiiHilo}%
    \and%
    Caltech/IPAC, 1200 E. California Blvd. Pasadena, CA 91125, USA\label{Caltech}%
    \and%
    Department of Computer Science, Aalto University, P.O. Box 15400, FI-00076 Espoo, Finland\label{DCS}%
    \and%
    Department of Physics, Faculty of Science, University of Helsinki, 00014 Helsinki, Finland\label{Helsinki}%
    \and%
    SOFIA Science Center, NASA Ames Research Center, Moffett Field CA 94035, USA\label{SOFIA}
    \and%
    Space Telescope Science Institute, 3700 San Martin Drive, Baltimore, MD 21218, USA\label{STScI}%
    \and%
    Kapteyn Astronomical Institute, University of Groningen, P.O. Box 800, 9700AV Groningen, The Netherlands\label{Groningen}
    \and%
    Purple Mountain Observatory, Chinese Academy of Sciences, 10 Yuanhua Road, Nanjing 210023, China\label{Nanjing}%
    \and%
    DTU Space, Technical University of Denmark, Elektrovej, Building 328, 2800, Kgs. Lyngby, Denmark\label{DTU}%
    \and%
    Department of Physics, University of California Santa Barbara, Santa Barbara, CA 93106, USA\label{SantaBarbara}%
    \and%
    European Space Agency (ESA), European Space Astronomy Centre (ESAC), Camino Bajo del Castillo s/n, 28692 Villanueva de la Cañada, Madrid, Spain\label{ESAC}%
    \and%
    Jet Propulsion Laboratory, California Institute of Technology, 4800 Oak Grove Drive, Pasadena, CA 91109\label{NASA}%
    \and%
    Department of Physics and Astronomy, UCLA, PAB 430 Portola Plaza, Box 951547, Los Angeles, CA 90095-1547\label{Losangeles}%
    \and%
    Department of Astronomy and Astrophysics, University of California, Santa Cruz, 1156 High Street, Santa Cruz, CA 95064 USA\label{UnivCalifornia}%
    \and%
    Institute for Astronomy, University of Hawaii, 2680 Woodlawn Drive, Honolulu, HI 96822, USA\label{HawaiiHonolulu}%
    \and%
    NASA Headquarters, 300 Hidden Figures Way, SE, Mary W. Jackson NASA HQ Building, Washington, DC 20546, USA\label{Jack-NASA}%
    \and%
    University of Bologna - Department of Physics and Astronomy “Augusto Righi” (DIFA), Via Gobetti 93/2, I-40129, Bologna, Italy\label{Bologna}%
    \and%
    INAF - Osservatorio di Astrofisica e Scienza dello Spazio, Via Gobetti 93/3, I-40129, Bologna, Italy\label{INAF}%
    \and%
    European Southern Observatory, Karl-Schwarzschild-Str. 2, D-85748, Garching bei München, Germany\label{ESO-Ger}%
    \and%
    Astronomy Centre, University of Sussex, Falmer, Brighton BN1 9QH, UK\label{Brighton}%
    \and%
    Department of Astronomy, University of Massachusetts, Amherst, MA 01003, USA\label{UMASS}%
   }

   \date{Received October 7, 2024; accepted March 11, 2025.}

 
  \abstract
   {The stellar mass-star formation rate ($\SFRM$) plane is an essential diagnostic to separate galaxy populations. However, we still lack a clear picture of how galaxies move within this plane along cosmic time.
  }
   {This study aims to provide an observational description of galaxy migrations in the $\SFRM$ plane based on the reconstructed star formation histories (SFH) of a sample of galaxies at redshift $z<4$. Ultimately, this study seeks to provide insight into physical processes driving star formation.}
   {We used data from the COSMOS field, which provides extensive multi-wavelength coverage. We selected a sample of 299131 galaxies at $z<4$ with the COSMOS-Web NIRCam data at a magnitude of $m_{\rm F444W}<27$ over a large area of 0.54 deg$^2$. We utilized the SED modeling code \cigale{}, which incorporates non-parametric SFHs, to derive the physical properties and reconstruct the SFHs of this galaxy sample. To characterize the SFHs and interpret the galaxies' movements on the $\SFRM$ plane, for each galaxy we also defined a migration vector in order to track the direction ($\Phidt$[deg]) and velocity norm ($\rdt\left[\rm{dex/Gyr}\right]$) of the evolutionary path over the $\SFRM$ plane. 
   We quantified the quality at which these migration vectors can be reconstructed using the {\sc Horizon-AGN} cosmological hydrodynamical simulation.
   }
   {We find that galaxies within the main sequence exhibit the lowest amplitude in their migration and a large dispersion in the direction of their movements. We interpret this result as galaxies oscillating within the galaxy main sequence. By using their migration vectors to find the position of main-sequence progenitors, we obtained that most of the progenitors were already on the main sequence as defined one billion years earlier. We find that galaxies within the starburst or passive region of the $\SFRM$ plane have very homogeneous properties in terms of recent SFH ($<$1 Gyr). Starburst galaxies assembled half of their stellar mass within the last 350\,Myr, and this population originates from the main sequence. Galaxies in the passive region of the plane show a homogeneous declining SFH over the full considered redshift range. We identified massive galaxies already in the passive region at $3.5<z<4$, and their number density increases continuously with cosmic time. The progenitors of passive galaxies are distributed over a large range of SFRs, with less than 20\% of passive galaxies being starburst 1\,Gyr earlier, thus shedding light on rapid quenching channels.}
   {
   Using reconstructed SFHs up to $z<4$, we propose a coherent picture of how galaxies migrate over cosmic time in the $\SFRM$ plane, highlighting the connection between major phases in the SFH.  
   }


   \keywords{galaxy evolution, astrostatistics, fundamental parameters, star formation }

   \authorrunning{Arango-Toro et al.}
   \titlerunning{Galaxy migrations in the $\SFRM$ plane}

   \maketitle
   \nolinenumbers

%
\section{Introduction}

The distribution of galaxies on the stellar mass–star formation rate plane represents the relationship between the star formation rate (SFR) - which quantifies the mass of new stars formed per year (measured in solar masses per year, [$\Msol , \text{yr}^{-1}$]) - and the stellar mass ($\Mstar$), which denotes the total mass of stars within a galaxy (measured in solar masses, [$\Msol$]). This representation provides crucial insights into the evolutionary state of galaxies. This plane is characterized by a tight correlation (with a scatter of 0.3 dex in SFR) between stellar mass and SFR for the majority of star-forming galaxies, often referred to as the "main sequence" (MS) of star-forming galaxies \citep[e.g.,][]{Brinchmann2004, Noeske07, Elbaz07, Daddi2007}, with the normalization of the MS increasing with redshift \citep[e.g.,][]{popesso23,Ciesla2024}. Deviations from this sequence can indicate different evolutionary pathways, such as quenching, where star formation is significantly reduced or halted, or starburst (SB) events, where star formation is temporarily enhanced \citep[see][]{Rodighiero2011,Rinaldi2022,Ciesla2023,Shi2024}.

Passive galaxies are located below the MS. They are often older and more massive and have lower SFRs \citep[e.g.,][]{peng2010,ilbert13,weaver22b}. Conversely, SB galaxies are located above the MS and exhibit intense and short-lived bursts of star formation that can drastically alter their properties \citep[e.g.,][]{Elbaz2011}. The $\SFRM$ plane also includes regions where galaxies could be transitioning from star-forming to quiescent states, often referred to as the "green valley" \citep[GV, e.g.,][]{Salim2014,Lin2023}. This intermediate region is critical for understanding the processes that regulate star formation and lead to quenching.

Star formation histories (SFHs) of galaxies provide a record of the star formation activity over cosmic time. The SFHs can be influenced by various mechanisms, including internal processes such as supernova feedback and active galactic nuclei (AGN) feedback as well as external processes such as mergers, tidal interactions, and environmental effects (e.g., ram-pressure stripping in clusters; \citep{Schawinski2014,Tacchella2016}). For instance, mergers can trigger bursts of star formation and fuel central AGN activity that lead to SB phases and subsequent quenching \citep[][]{Hopkins2008,Stone2024}. Alternatively, environmental effects can remove gas from galaxies and prevent the accretion of new incoming gas from the cosmic web, thus suppressing star formation and accelerating quenching \citep[][]{Dekel2008,Peng2015,Lofaro2024,Das2024}.

Ideally, SFHs are primarily constrained through spectroscopy \citep[e.g.,][]{Goddard17,Chauke18,Schreiber18,Steel24,IglesiasNavarro24}.
Indeed, the best indicators allowing for their reconstruction are Balmer emission and absorption lines. The H$\alpha$ line is a direct probe of the star-formation activity in galaxies, and it is sensitive to very short timescales ($\sim$10\,Myr). When associated with the Balmer break, which probes the older stellar component, they strongly constrain the SFHs of galaxies. However, these measurements are limited to either large samples of local galaxies or small samples of individual galaxies at higher redshifts.

Recent advancements in observational surveys and spectral energy distribution (SED) modeling have greatly enhanced the ability to study these processes from photometry. With its extensive multi-wavelength coverage, the COSMOS field provides a rich dataset for analyzing galaxy properties across different cosmic epochs \citep{scoville07,Capak2007,laigle16}. In particular, the recent COSMOS-Web survey, which leverages the capabilities of the James Webb Space Telescope, offers unprecedented depth and resolution in the near-infrared (NIR) over the COSMOS field, facilitating the study of faint and distant galaxy populations \citep{casey23}. The survey field of view of 0.54\,deg$^2$ enables detailed studies of the $\SFRM$ plane in different environments.

In parallel, developments of sophisticated SED fitting techniques improve the determination of physical parameters and SFHs. Early SED fitting codes such as Hyperz \citep{Bolzonella00}, Z-PEG \citep{LeBorgne2002}, \lephare{} \citep{Arnouts02,Ilbert06}, and MAGPHYS \citep{DaCunha08} laid the groundwork for multi-wavelength photometric redshift estimation and stellar population analysis. The latest generation of SED fitting codes, such as BEAGLE \citep{Chevallard16} or STARDUST \citep{Kokorev2021}, incorporate far-infrared data to account for dust attenuation and emission, thus improving physical parameter estimates and allowing for the reconstruction of complex SFHs by including recent quenching or burst events \citep{Ciesla17,Schreiber18,Aufort20}. Moreover, advanced SED fitting codes such as BAGPIPES \citep{Carnall2018}, PROSPECTOR \citep{Johnson2021,leja17}, or \cigale{} \citep{Boquien19} now include non-parametric SFH modeling approaches \citep{Leja19,Tacchella2021,Ciesla23}.

Historically, SFH reconstructions often relied on simple analytical models, such as exponentially declining models, which could introduce biases and fail to capture the complexities of star-formation activities \citep[e.g][]{Pacifici12,Simha14}. These models often assumed smooth monotonically decreasing SFHs, which do not accurately reflect the stochastic nature of star formation, particularly in galaxies experiencing bursts or quenching \citep{Papovich01,Kauffmann03}. Intermediate analytical solutions incorporated more flexibility by allowing for additional parameters to model bursts of star formation or quenching events, leading to improved fits to observed data \citep{Dressler04,Gladders2013}. However, these models still imposed constraints that could limit their ability to accurately represent the true SFH.

The advent of non-parametric SFH models represents a significant advancement, as they allow for more detailed and unbiased reconstructions of SFHs. These models do not assume a specific functional form for the SFH but instead use a series of time bins with independently determined SFRs, thus providing a more flexible and realistic depiction of the star-formation activities over cosmic time \citep{Tojeiro09,Iyer19}. Studies utilizing non-parametric approaches, such as those by \citet{Leja19}, \citet{Pacifici2023}, and \citet{Ciesla23}, have demonstrated the ability to recover physical parameters and complex SFHs, including SFH features such as multiple bursts and periods of quiescence, that parametric models often miss. This flexibility allows for a more accurate understanding of the processes driving galaxy evolution and highlights the importance of using sophisticated models to interpret the wealth of data from modern surveys and large statistical samples. Especially important in this context is the ability to capture star formation burstiness, which seems to be the most likely explanation for many JWST high-redshift observed galaxies \citep[e.g.,][]{Sun2023}.

The position of a galaxy in the $\SFRM$ plane is commonly used to separate galaxy populations based on their observable properties. However, it is not clear if galaxies at a given position of the $\SFRM$ plane represent a homogeneous galaxy population in terms of SFHs or if galaxies have experienced a large variety of histories \citep[e.g.,][]{Elbaz2018,Ciesla23}. Furthermore, the characterization of the MS and its evolution has vastly improved over the past decade \citep[e.g.,][and references therein]{popesso23}, providing detailed insights into galaxy properties at various epochs. It is still difficult to understand how and when galaxies transition from different regions of the $\SFRM$ plane. These movements should be encoded in the SFH of galaxies \citep[e.g.,][]{Tacchella2016}. In this work, we combine state-of-the-art SED modeling methods and the rich dataset of the COSMOS-Web survey to characterize and analyze the movement of galaxies with the $\SFRM$ plane by reconstructing their SFH. 

This work is organized as follows. Section~\ref{Sec:survey} describes the COSMOS-Web survey and the photometric data employed for our analysis. Section~\ref{sec:cigale} presents the code investigating galaxy emission, \cigale,\,  which has been used to recover the physical parameters. In Sect.~\ref{sec:AGN}, we test the reliability of our SED fitting approach to recover physical parameters as well as the SFH from the mock photometry of the {\sc Horizon-AGN}~\citep{dubois14} cosmological hydrodynamical simulation. Section \ref{sec:sample} outlines the construction of our observational data set from the COSMOS-Web survey. We analyze galaxy migration in the $\SFRM$ plane in Sect.~\ref{sec:results}, and this is followed by a discussion in Sect.~\ref{sec:disc}.

We adopted the standard  $\Lambda$CDM cosmology with $\Omega_\textrm{m}=0.3$, $\Omega_\Lambda=0.7$, and $H_0=70\,\rm km\,s^{-1}\,Mpc^{-1}$. We used the initial mass function (IMF) from \cite{Chabrier03}. The magnitudes are given in the AB system \citep{oke_absolute_1974}.

\section{The COSMOS-Web survey \label{Sec:survey}}

\subsection{The imaging data and photometric catalog}\label{Subsec:catalog}

The COSMOS-Web survey \citep[PIs: Jeyhan Kartaltepe and Caitlin Casey]{casey23} is a 255-hour Cycle 1 observation program conducted with the James Webb Space Telescope. This survey covers 0.54\,deg$^2$ in four NIRcam filters \citep{rieke23} reaching a 5$\sigma$ point-source depth between $\sim 27.5-28.2$ mag. The field is covered by 152 visits in each band, which are used to create the NIRCam mosaics centered at $\alpha=$10:00:27.92,
$\delta=$+02:12:03.5. The observations were conducted using the F115W, F150W, F277W, and F444W filters (see \citealp{casey23} for more details). In parallel, the survey covers a non-contiguous 0.19\,deg$^2$ area imaged in a single MIRI filter (F770W) \citep{wright22} reaching a 5$\sigma$ point source depth between $\sim 25.3-26.0$\,mag.

The multi-wavelength legacy of the COSMOS field \citep{scoville07,Capak2007} offers a wide and deep coverage from X-rays \citep{civano16,marchesi16} to radio wavelengths \citep{schinnerer10,smolcic17}.
In the optical regime, the available photometric measurements of the COSMOS-Web survey include data in the $u$ band with MegaCam at the Canada-France-Hawaii telescope (CFHT) reaching a depth of $\sim 27.0$\,mag \citep{sawicki19}; the Advanced Camera for Surveys (ACS) data from the Hubble Space Telescope (HST) in the F814W band with high-resolution imaging covering 1.64 deg$^2$ of the whole COSMOS field \citep{koekemoer07} reaching a 5$\sigma$ point-source of $\sim 27.2$\,mag; the Hyper Suprime-Cam (HSC) imaging in the $g,~r,~i~,z$ and $y$ bands, deeper than $\sim 26.5$\,mag, including also imaging in eleven intermediate-bands and two narrow bands from the Subaru Suprime-Cam between 4266\,\AA{} and 8243\,\AA{} \citep{taniguchi15}. In the near-infrared regime, we include the ground-based UltraVISTA survey \citep{McCracken12,Moneti23} in the $Y,~J, ~H$ and $K_{\rm s}$ bands between 1 and 2.2\,$\mu$m to complement the NIRCam and MIRI imaging. We refer the reader to table~\ref{tab:band_infos}, which resumes the different bands used in this work.

Given the new NIRCam observations, a photometric catalog specific to the COSMOS-Web survey has been developed, which supersede previous COSMOS photometric catalogs \citep{ilbert13,laigle16,weaver22b}.
The sources are detected over a chi-square ( $\chi^2$) detection image combining PSF-homogenized images in the four NIRCam filters. 
The high resolution of NIRCam images allows us to separate sources previously blended when the detection was performed on ground-based images \citep{weaver22b}. We detect more than 784000 sources over 0.54 deg$^2$. Conventional approaches to measuring photometry in fixed apertures are impractical due to the wide range of PSF sizes going from $0\overset{\arcsec}{.}05$ to 1\arcsec{} in space and ground-based datasets. We used \texttt{SourceXtractor++} \citep{SX++22} to model Sersic surface brightness profiles \citep{sersic} in native resolution NIRCam imaging, then extract photometry using that model in each band considering their own PSF. When several sources potentially overlap in the ground-based data, they are grouped and fit simultaneously. We refer the reader to Shuntov et al., in prep for a detailed description of the COSMOS-Web photometric catalog.

\subsection{Photometric redshifts}

The photometric redshifts used for the COSMOS-Web catalog are extensively described in Shuntov et al., in prep as well as the accuracy of the \photoz{} measurements. We summarize the main features in this section.

The photometric redshifts are computed using the template-fitting code \lephare{}
\citep{Arnouts02, Ilbert06}. Given the deep
NIRCam coverage, the survey includes a large variety of
galaxies. The deep and extensive multi-wavelength coverage
available in COSMOS allows us to enlarge the parameter space probed by our
template library compared to previous COSMOS catalogs
\citep[e.g.,][]{ilbert13,laigle16,weaver22b}. We based the template library on $12$ templates
generated with the \citet[][hereafter BC03]{BruzualCharlot03}  
stellar population synthesis models, at 42 different ages. These
templates are described in \citet{ilbert15}. They are
created assuming different SFHs (exponentially
declining and delayed) and two different metallicities  ($Z=0.008Z_\odot$,
$0.02Z_\odot$). Emission lines are included following a method similar to
\citet{schaerer09}, and we allowed the line
intensity to vary by a factor of two. The dust attenuation is added
as a free parameter with $E(B-V)$ varying from $0$ to $1$,
considering three possible attenuation laws
\citep{calzetti2000,arnouts13,salim18}. The energy absorbed
in ultraviolet (UV)-to-optical range is assumed to be fully remitted in IR by the dust. We model the dust emission by using a template library from  \citet{Bethermin12} based on \citet{magdis12}; the dust template is rescaled to the expected IR luminosity using energy balance. Inter-galactic medium absorption is implemented following the prescription from \citet{madau95}.

The photometric redshifts are computed on the \texttt{SourceXtractor++} model photometry described in the previous section, and the bands are listed in Table  \ref{tab:band_infos}. As described in Shuntov et al., in prep, we rejected sources falling in masked areas (e.g., bright halos close to bright stars) or classified them as potentially spurious. The \photoz{} accuracy is assessed by comparing their values with a compilation of spectroscopic redshifts (hereafter \specz{}) established by Kostovan et al., in prep. This compilation includes more than  11000 \specz{} with a confidence level larger than 97\% 
\citep{lilly07,lilly09,kartaltepe10,kartaltepe15b,silverman15,kashino19,lefevre15,casey12,capak11,kriek15,hasinger18}. The precision in the \photoz{} ($z_{\rm p}$) depends primarily on the considered magnitude: $\sigmaz{}<$ 0.01 at $m_{\rm F444W}<24$ and 2\% of catastrophic failure (defined as the $z_{\rm p}$ for which $\left|z_{\rm p}-z_{\rm s}\right|>0.15\times(1+z_{\rm s})$ with $z_{\rm s}$ as the \specz{}). Thanks to the extremely deep NIRCam coverage, the \photoz{} precision remains lower than 0.03 at  $26<m_{\rm F444W}<27$ and $z<4$ with 15\% of failures, which are the faintest magnitudes and highest redshifts considered in this study. 

As a final note, \lephare{} is optimized to derive high-quality photometric redshifts with several specific options, tested extensively in COSMOS \citep[][]{Ilbert09, ilbert13,laigle16, weaver22b}. However, the code does not include non-parametric SFH explaining the need for using \cigale{} in the next section.

\section{SED modeling with \cigale\ } \label{sec:cigale}

\begin{table}
\caption{Ultraviolet-optical-IR data used in the SED fitting}
\footnotesize
\setlength{\tabcolsep}{2pt}
\begin{threeparttable}

\begin{tabular*}{\hsize}{lcccc}
 \hline \hline
Instrument & Band & Central\tnote{a} & Width\tnote{b} & Depth\tnote{c} \\
/Telescope &  & $\lambda$ [\AA{}] & [\AA{}] & (1.0\arcsec, 0.15\arcsec, 0.5\arcsec)\\
(Survey) &  &  &  &  \\
 \hline
MegaCam & $u$ & 3858 & 598 & 27.3  \\
/CFHT  \\
 \hline
ACS/HST & F814W & 8333 & 2511 & 27.5 \\
 \hline
HSC & $g$ & 4847 & 1383 & 27.6 \\
/Subaru & $r$ & 6219 & 1547 & 27.2  \\
HSC-SSP & $i$ & 7699 & 1471 & 27.0  \\
PDR3 & $z$ & 8894 & 766 & 26.6 \\
 & $y$ & 9761 & 786 & 26.0  \\
 \hline
Suprime-Cam & IB$427$ & 4266 & 207 & 25.7  \\
/Subaru & IA$484$ & 4851 & 229 & 26.2  \\
 & IB$505$ & 5064 & 231 & 25.9  \\
 & IA$527$ & 5261 & 243 & 26.1  \\
 & IB$574$ & 5766 & 273 & 25.3  \\
 & IB$624$ & 6232 & 300 & 26.1 \\
 & IB$679$ & 6780 & 336 & 25.3  \\
 & IB$709$ & 7073 & 316 & 25.7  \\
 & IB$738$ & 7361 & 324 & 25.8  \\
 & IB$767$ & 7694 & 365 & 25.3  \\
 & IB$827$ & 8243 & 343 & 25.3  \\

 \hline
VIRCAM & $Y$ & 10216 & 923 & 25.8  \\
/VISTA & $J$ & 12525 & 1718 & 25.8  \\
UltraVISTA & $H$ & 16466 & 2905 & 25.5  \\
DR5  & $K_s$ & 21557 & 3074 & 25.3  \\
\hline
NIRCam & F115W & 11540 & 2250 & 27.2  \\
 & F150W & 15010 & 3170 & 27.4  \\
 & F277W & 27760 & 6730 & 28.1  \\
 & F444W & 44010 & 10230 & 28.0 \\
 \hline
 MIRI & F770W & 76390 & 19500 & 25.2  \\
 \hline
\end{tabular*}
\begin{tablenotes}
\item[a] Median of the transmission curve.
\item[b] Full width of the transmission curve at half maximum.
\item[c] $5\sigma$ depth computed in empty apertures with diameters of 1.0\arcsec~for the ground-based, 0.15\arcsec~for the space-based JWST/NIRCam and HST/ACS and 0.5\arcsec~for JWST/MIRI images, averaged over the NIRCam area.
\end{tablenotes}
\end{threeparttable}

\label{tab:band_infos}
\end{table}

In addition to the SED fitting done with \lephare{} to get the photometric redshifts, we also used the SED modeling code \cigale\footnote{\url{https://cigale.lam.fr/}} \citep{Boquien19} to derive the physical properties of galaxies and their SFHs. \cigale\ builds and fits physical models spanning from X-ray to radio wavelengths, accounting for the energy budget between the light absorbed in the UV-optical range by dust and reemitted in IR. It performs a Bayesian-like analysis to derive the physical parameters of galaxies. Its versatility is characterized by the multiple modules that model the galaxy SFH, the stellar, dust, and nebular emission, the AGN contribution, and the radio emission of galaxies. 
In particular, the SFH can be handled through analytic, non-parametric, or simulated models \citep{Boquien14, Ciesla15, Ciesla17, Ciesla23}. We used the recently added non-parametric SFH module \texttt{sfhNlevels} \citep{Ciesla23}, BC03 stellar population models, the \citet{Dale14} dust emission library and a modified \citet{calzetti2000} attenuation law, which implements a modified SB law with the continuum attenuated with a Calzetti (2000) curve and the lines extinction with a Milky Way or a Magellanic Cloud attenuation curve.
Table~\ref{cig_tab} lists the parameters used in the \cigale\ SED fitting, where a Chabrier IMF was assumed with two stellar metallicities, sub-solar and solar values (Z = 0.008, 0.02). 

A probability distribution function is associated with each parameter estimated by CIGALE. The flux uncertainties are included in the estimated $\chi^2$. In addition, a 10\% uncertainty is added in quadrature to the flux to account for model uncertainties, as implemented in \cigale\ \citep[see][]{Boquien19}.

\cigale{} is applied to the photometric catalog described in Sect.~\ref{Subsec:catalog} (i.e. neither X-ray, far-IR, or radio data are considered in the fit). The redshift is set to the value derived previously by \lephare{}. The comparison between the stellar masses derived by the two codes shows a dispersion of 0.12 dex with a \cigale{}'s systematic bias of +0.17 dex. However, the dispersion between the SFR derived by the two codes reaches 0.4 dex with a bias of -0.15 dex on \cigale{}'s SFR measurements. This difference could be partially explained by the use of non-parametric SFH adopted in \cigale{} and different attenuation laws.

\begin{table}
\caption{\cigale\ Spectral energy distribution fitting models and templates.}
\footnotesize
\setlength{\tabcolsep}{1pt}
    \begin{threeparttable}
    \begin{tabular*}{\hsize}{lll}
    \hline
    \hline
    \multicolumn{3}{c}{Star formation history}\\
    \hline
    \multicolumn{3}{c}{Non-parametric: \texttt{sfhNlevels (1)}}\\ 
    \hdashline
    $age$\tnote{a}{[}Myr{]} & {[}\small1578;13753{]} & \small{Age of the oldest stars in the galaxy in Myr;}\\ 
    &&\small{100 values linearly sampled.}\\
    1$^{\rm{st}}$bin\tnote{b}[Myr]  &     10 & \small{Duration of the first bin in lookback time.}\\
    N$_{\rm bins}\tnote{c}$&10&\small{Number of bins in the SFH.}\\
    N$_{\rm SFH}$\tnote{d}& 2000 &\small{Number of SFHs drawn for each $age$ value.}\\
    \hline
    \hline
    \multicolumn{3}{c}{Simple stellar population: \texttt{BC03} (2) }\\
    \hdashline
    IMF & Chabrier&Initial Mass Function\\
    $Z$ & 0.008,0.02&Metallicity\\
    \hline
    \hline
    \multicolumn{3}{c}{Attenuation: \texttt{dust\_att\_modified\_starburst} (3)}\\ 
    
    \hdashline
    $E(B-V)s$ & [0,1.8] & \small{Color excess; 10 values linearly sampled}\\ 
    \hline
    \hline
    \multicolumn{3}{c}{Dust emission: \texttt{dale2014} (4)}\\ 
    \hdashline
    $\alpha$ &  2.0  &  \small{Far-IR slope}\\ 
    
    \hline
    \end{tabular*}
\tablebib{(1)~\citet{Ciesla22, Ciesla23}; (2)~\citet{BruzualCharlot03}; (3) \citet{calzetti2000}; (4) \citet{Dale14}}
\begin{tablenotes}
\item[a] This module selects an $age$ from a linearly sampled range of 100 values between 1578 and 13753 Myr, which corresponds to the age of the Universe at the redshift of the galaxy. 
\item[b] The size of the first bin is set at 10 Myr. 
\item[c] N$_{\rm bins}$-1 logarithmic spaced time-bins are computed between 10 Myr and the selected $age$.
\item[d] For each sampled $age$ value, 2000 SFHs are drawn, with the SFR amplitudes of each time-bin computed following a student-t distribution with continuity-burst priors \citep{Leja19,Tacchella22}.
\end{tablenotes}
\end{threeparttable}
\label{cig_tab}
\end{table}

\subsection{\label{non-param} Non-parametric SFH: The sfhNlevels module}

The \cigale\ non-parametric ``sfhNlevels'' module models the SFH using a given number of bins with a constant SFR, where the SFRs of two consecutive bins are compelled by a given prior. In this work, we infer the piece-wise constant SFH with a Bayesian analysis of the photometry using a continuity-burst prior that penalizes sharp variations of the SFH with a Student-t distribution \citep{Leja19,Tacchella22}. The number of bins describing the SFHs (N$_{\textrm{bins}}$) and the age of the first bin in lookback time ($1^{\textrm{st}}~\textrm{bin}$) are set as fixed parameters as well (see Table~\ref{cig_tab} for details). The SFH covers the age of the Universe at the redshift of the considered galaxy. The choice of 2000 SFHs drawn from the prior (N$_{\rm{SFH}}$), after fixing the redshift, ensures sufficient coverage of the posterior while balancing computational efficiency. To confirm this choice, we conducted a dedicated test using the SED fitting framework by varying the number of SFH draws: 100, 500, 1000, 2000, 4000, and 5000. We examined the coverage of the NUV-r-K color-color space for each case and found that with SFH draws $\geqslant$ 2000, the color-color space is well sampled, indicating adequate posterior coverage. 
Furthermore, this configuration generates over 400 million models when combined with other free parameters, ensuring the feasibility of the SED fitting procedure within a computationally intensive workflow. The choice of the continuity-burst priors for our galaxy dataset is motivated by several key factors. Galaxies at redshifts higher than 2 often experience significant episodic star formation events, such as bursts triggered by mergers or interactions \citep{Haskell24}. The continuity-burst prior allows for low to moderate sharp changes in the SFR which are not well-represented by smoother continuity priors \citep{park24}. This flexibility is crucial for accurately capturing the SFHs of dynamically evolving galaxies undergoing various phases, including SB periods followed by quiescent phases \citep{Rowlands18}. This behavior is particularly prevalent in the early universe due to the higher interaction rates and the availability of cold gas \citep{Madau14}. The continuity-burst priors model these phases more effectively, proving a more realistic reconstruction of SFHs across a broad range of redshifts \citep{Tosi09}.

\begin{figure}[H]
\centering
    \includegraphics[width=0.8\columnwidth]{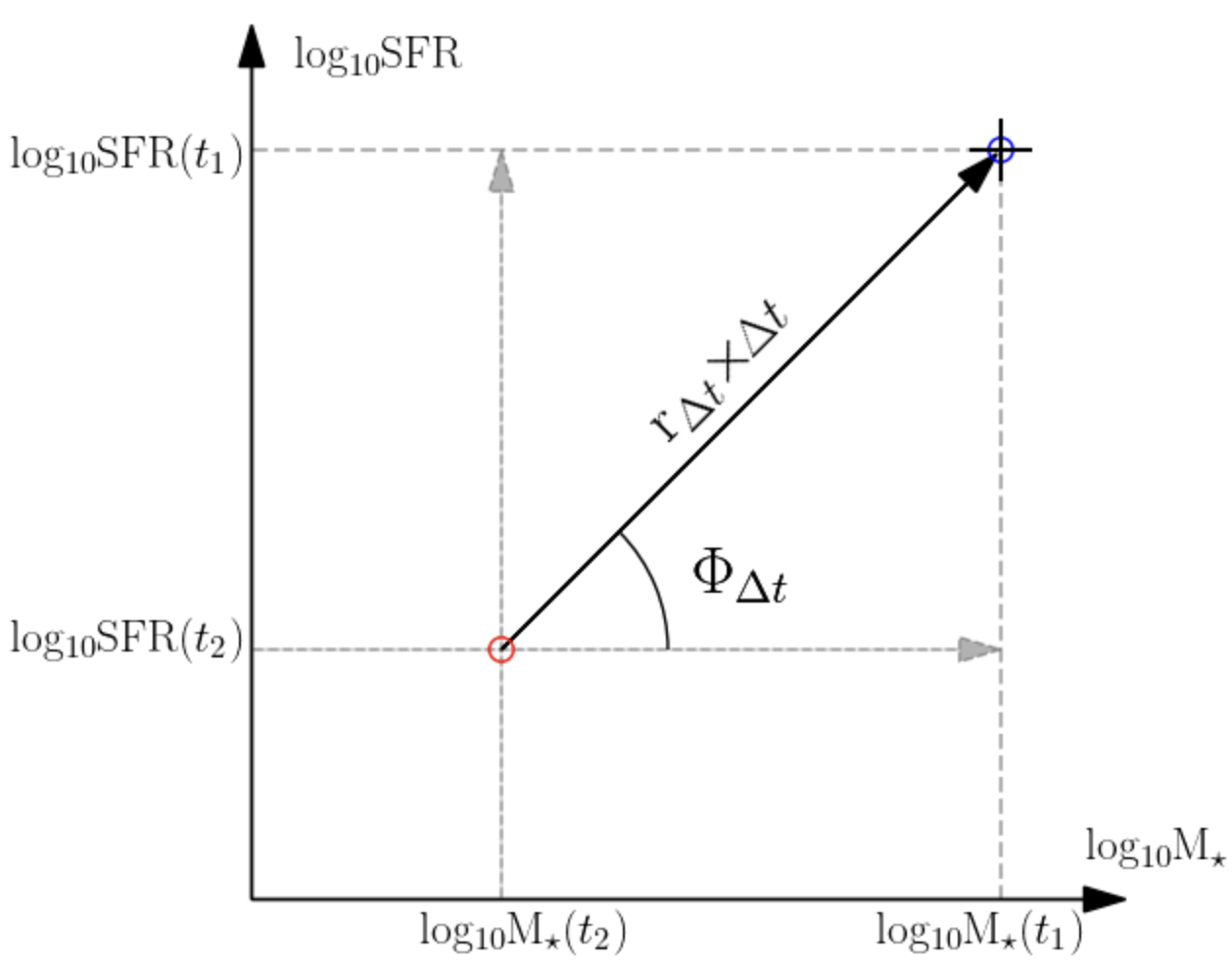}
       \caption{Illustration of the migration vector definition ($\migvec$). The output SFH from CIGALE provides the SFRs and stellar masses ($\Mstar$) of the galaxy between the times t$_2$ and t$_1$, allowing its placement on the $\SFRM$ plane at that epoch. The migration vector is defined by the angle $\Phidt$ (Eq. \ref{grad_sfr}) between the line connecting the galaxy's position and the horizontal line representing a constant SFR. The quantity $ r_{\Delta t} \times \Delta t$ (Eq. \ref{norma}) represents the norm of the vector connecting the galaxy's position, where $ r_{\Delta t}$ is the rate of change of the galaxy's position on the $\SFRM$ plane over time interval $ \Delta t $}
       \label{illust}
\end{figure}

\begin{figure}[]
    \includegraphics[width=\columnwidth]{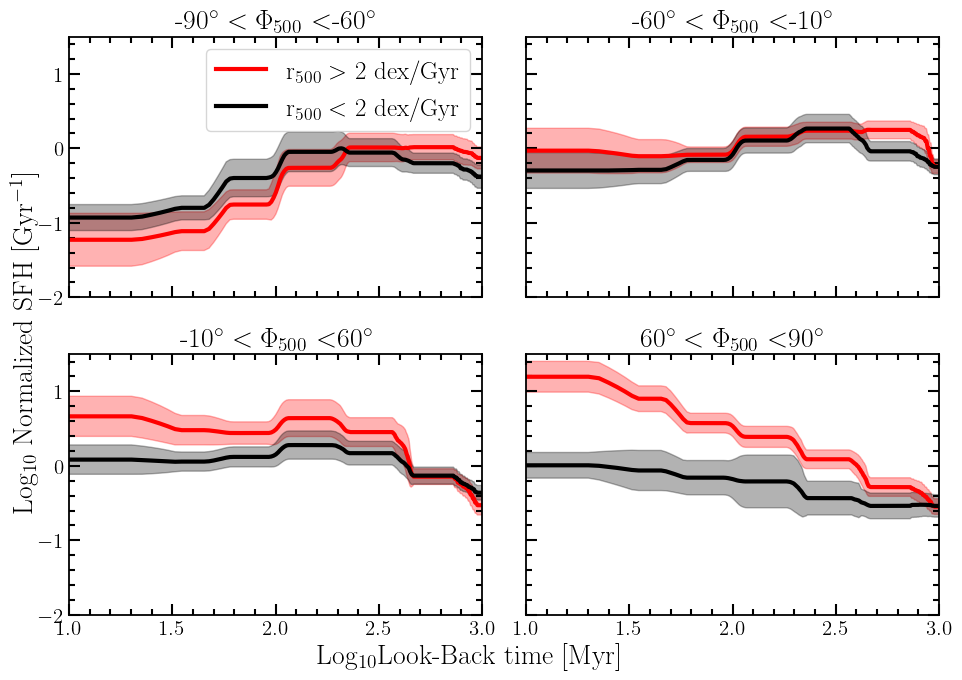}
       \caption{Posterior mean SFHs derived for the COSMOS-Web galaxies via SED modeling. Galaxies are classified based on their migration angle ($\Phi_{\Delta t}$) over the last 500\,Myr. Galaxies with $\Phi_{\Delta t}$ between $-90^\circ$ and $-60^\circ$ exhibit fast-declining SFH, while those with $\Phi_{\Delta t}$ ranging from $-60^\circ$ to $-10^\circ$ show slow decline. Star-forming galaxies with relatively recent flattening of their SFH fall within $\Phi_{\Delta t}$ between $-10^\circ$ and $60^\circ$. In contrast, those with constantly rising recent SFH are categorized with $\Phi_{\Delta t}$ between $60^\circ$ and $90^\circ$. These classifications highlight diverse evolutionary trajectories among galaxies. The red and black lines correspond to selected galaxies with higher and lower migration vector norms to the median values ($\rfive=2$dex/Gyr), respectively. Shaded regions present the confidence intervals.}
       \label{grad_sfh}
\end{figure}

For this particular work, we have included the computation of a vector that characterizes the migration of a galaxy over the $\SFRM$ plane in a lookback time interval $\Delta t= t_2-t_1$, with $t_1$ being the lookback time at the redshift of the considered galaxy and $t_2$ being a lookback time with $t_2>t_1$. Hereafter, this vector is called migration vector and denoted $\migvec$. We characterize this vector by its angle, measured from the projected $x$-axis (noted $\Phidt$) and its velocity norm (noted  $\rdt$). These quantities are defined as 
\begin{equation}
    \Phidt = \arctan \left(\frac{\Delta \textrm{log}_{10}\textrm{SFR}_t}{\Delta \textrm{log}_{10}\textrm{M}_{\star t}}\right),
    \label{grad_sfr}
\end{equation}
\begin{equation}
    \rdt^2= \frac{(\Delta \textrm{log}_{10}\textrm{SFR}_t)^2+(\Delta \textrm{log}_{10}\textrm{M}_{\star t})^2}{\Delta t ^2}.
    \label{norma}
\end{equation}
They measure the direction (in degrees) and the velocity norm (in dex/Gyr) in a given lookback time interval $\Delta t$ in which the SFR changes by $\Delta \textrm{log}_{10}$SFR$_{t}$ and the stellar mass by $\Delta \textrm{log}_{10}$M$_{\star t}$ (without taking in account the returning fraction for this measurement\footnote{In this work, we adopt the term stellar mass to refer specifically to the mass currently residing in living stars and stellar remnants. In contrast, total mass formed refers to the total mass of stars that have formed throughout the galaxy's history, including both the currently existing stars and those that have already evolved into stellar remnants. It is important to make this distinction to avoid any potential confusion, as the term stellar mass is sometimes used in the literature to refer solely to the present-day mass in stars, without accounting for the mass formed earlier in the galaxy's evolutionary timeline.}, see \citealp{Leitner2011}). We refer the reader to Fig.~\ref{illust} for a schematic illustration of the migration vector.
With this definition, galaxies with $-90^\circ<\Phidt<0^\circ$ will roughly present a declining star formation activity, and galaxies with $0^\circ<\Phidt<90^\circ$ have an enhanced star formation activity over the last time interval $\Delta t$. For a constant SFR, $\Phidt=0$. As an illustration, we show in Fig.~\ref{grad_sfh} examples of SFHs selected from our COSMOS-Web sample, set by different values of $\Phidt$ and $\rdt$ over $\Delta t=500$\,Myr (we adopt this value arbitrarily for illustration purposes within the considered time range explored in this paper).
This modified module of \cigale\ and the $\SFRM$ migration velocity vector constitutes an updated version of the \cigale\ \texttt{sfhNlevels} module already defined and tested in \citet{Ciesla22, Ciesla23}, and \citet{Arango23}.

\subsection{Mock analysis}\label{sec:mock}

\begin{figure*}
     \includegraphics[width=1\textwidth]{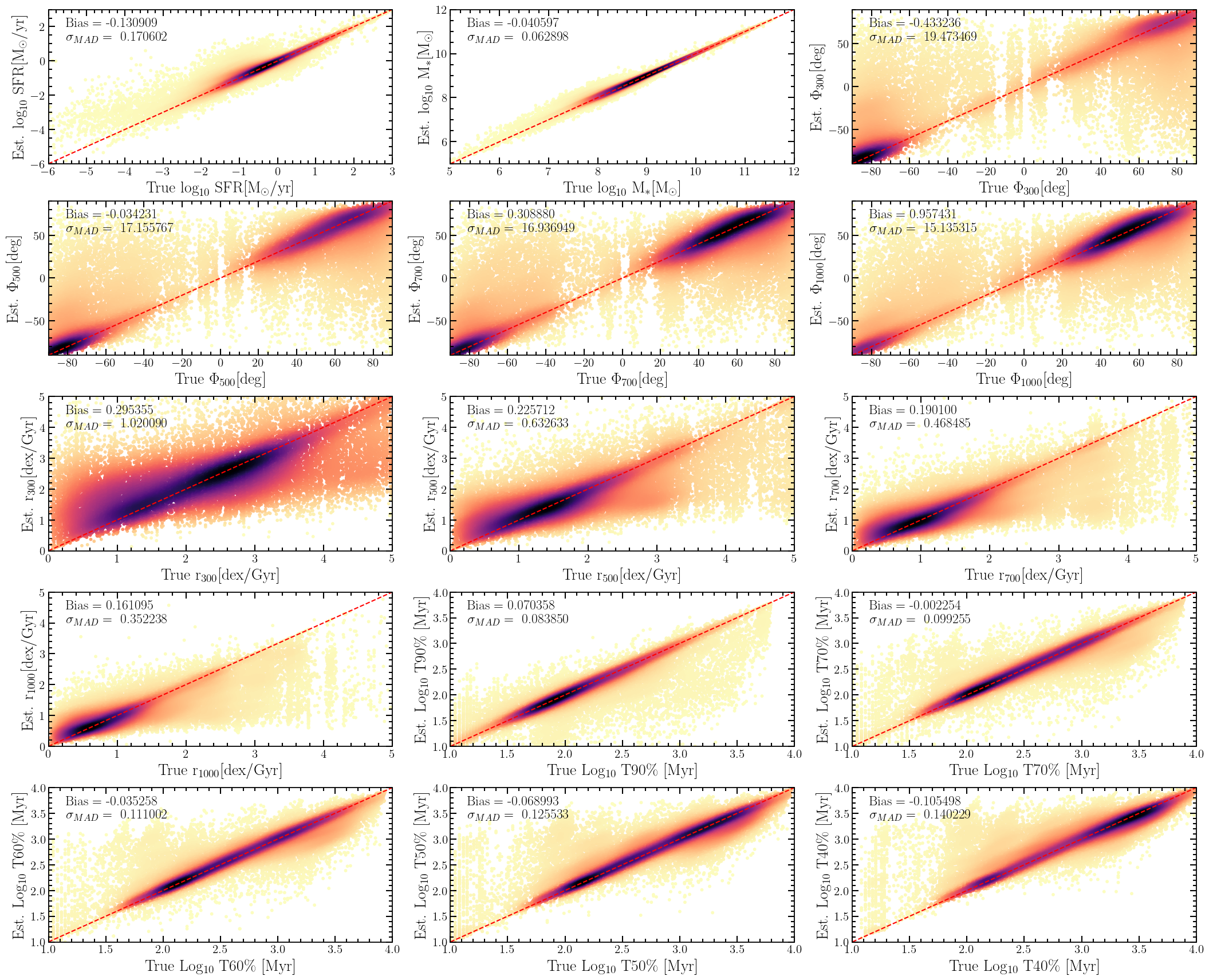}
        \caption{Results of the mock analysis on the SED modeling using the non-parametric SFH model. The input parameters used to build the mock catalog are shown on the $x$-axis, and the results from fitting the mock catalogs are shown on the $y$-axis. From left to right, the upper panels present the instantaneous SFR (here defined as the  SFH in the first bin of 10\,Myr), and the stellar mass $\Mstar$. The $\Phidt$ values are presented as defined on Eq.~\ref{grad_sfr} and computed over the last 300 to 1000\,Myr, on the uppermost right panel and over the second-row panels. The third row and the left panel of the fourth row count for $\rdt$ (i.e., the norm of the migration vector) computed over the same timescales as $\Phidt$. The two rightmost panels on the fourth row and bottom panels present the mass formation times (noted T$i$\%[Myr] and defined as the time taken to form the $\{i\}$ percentage of a galaxy’s total stellar mass) for the 90 to the 40\%. The solid black lines indicate the one-to-one relation. The bias and precision estimated for the parameters are indicated in each panel.}\label{mock_N-levels}
\end{figure*}

To assess the reliability of the parameters derived with \cigale, given the noise in the observations, we built a mock catalog mimicking the COSMOS-Web data sample, following the approach of \cite{Noll09,Buat14,Ciesla15}, and \citet{Boquien19}. For each galaxy of this  COSMOS-Web sample at $z<4$, we integrate the best-fit template obtained from a first \cigale\ run on the same set of filters as the original catalog. Then, these mock flux densities are perturbed by adding a randomly selected noise from a Gaussian distribution, with the Gaussian standard deviation $\sigma$ corresponding to the observed photometric error associated with the studied galaxy in this filter. We then run the code on this synthetic sample where all the parameters are known to compare the true and estimated synthetic values of the properties. We stress that we applied the same CIGALE configuration as used for the data, including the 10\% uncertainties in the simulated photometry. Figure~\ref{mock_N-levels} shows the results of the mock analysis for our COSMOS-Web galaxy sample. We find a good agreement for both the stellar mass and the SFR. The absolute bias on the instantaneous SFR (here defined as the SFH in the first bin of 10\,Myr) and the stellar mass is lower than 0.14 and  0.04\,dex, respectively. The dispersion (measured via the median absolute deviation $\sigma_{\textrm{MAD}}$) is relatively low for the stellar mass ($\sim0.06$\,dex) and remains lower than $\sim0.18$\,dex on the SFR indicator.

For this study, we analyze the results of mock analyses of several parameters directly measured from the SFHs. Specifically, the $\Phidt$ and $\rdt$ for several $\Delta t$ values. 
We find a $\sigma_{\textrm{MAD}}$ between $16^\circ$ and $20^\circ$ for $\Phidt$ measured with $\Delta t=1000$ and 300\,Myr, respectively. We find an absolute bias lower than 1$^\circ$, but this is likely due to the sign of the bias changing throughout the distribution. The bias is positive for negative $\Phidt$ values and negative for positive $\Phidt$ values, leading to a vanishing effect when the bias is measured over the full angle range. When measuring the bias as a function of $\Phidt$, the bias remains confined to the range of $-20^\circ \lesssim \text{bias}(\Phidt) \lesssim 20^\circ$. Therefore, a measured negative (positive) $\Phidt$ could be even more negative (positive) if this bias were corrected. We note a bimodality in the values of $\Phidt$, but this bimodality is present by construction in the COSMOS-Web data with a low fraction of SFH with a constant SFR or slowly declining SFR. $\rdt$ presents an absolute bias lower than 0.3\,dex/Gyr and relatively low $\sigma_{\textrm{MAD}}$ going from 1 to 0.4 dex/Gyr for $\Delta t$ measured with 300 and 1000\,Myr, respectively. This observed variation on the $\sigma_{\textrm{MAD}}$ for the $\rdt$ parameter may be attributed to the inherent stochasticity of recent SFHs or even the time resolution, given the broad-band photometric coverage of our data set \citep{Zibetti2024}. 

Furthermore, to examine the robustness of the derived SFHs, we compare different mass formation times, defined here as the lookback time at which a galaxy has formed a given percentage of a galaxy's total stellar mass (e.g., 0 Gyr for 100\%). This comparison allows us to evaluate the performance of \cigale{} in recovering both the shapes and amplitudes of galaxy SFHs, considering the uncertainty in the photometric measurements of the COSMOS-Web dataset. This analysis is illustrated in Fig.~\ref{mock_N-levels}, where the panels labeled true and estimates log$_{10}$T$i$\%[Myr], with $i$ ranging from 90\% to 40\% of the total formed mass along the SFH of galaxies, and present an absolute bias lower than $\sim 0.11$\,dex and a $\sigma_{\textrm{MAD}}\sim0.14$\,dex. Based on the results of this mock analysis, we have obtained reliable measurements of several key parameters, including SFR, M$_*$, and $\Phidt$ and $\rdt$, for $\Delta t$ ranging from 300 to 1000\,Myr. Additionally, the inferred SFHs provide evidence of star formation accounting for at least $\gtrsim$ 40\% of the total mass. However, it is important to note that these findings depend on the assumption that our models perfectly reflect the true universe. The next section partially alleviates such an assumption by using a more complex mock catalog based on cosmological simulations.

\section{Validating derived galaxy parameters through SED fitting and the Horizon-AGN simulation\label{sec:AGN}}

\begin{figure*}
    \includegraphics[width=1\textwidth]{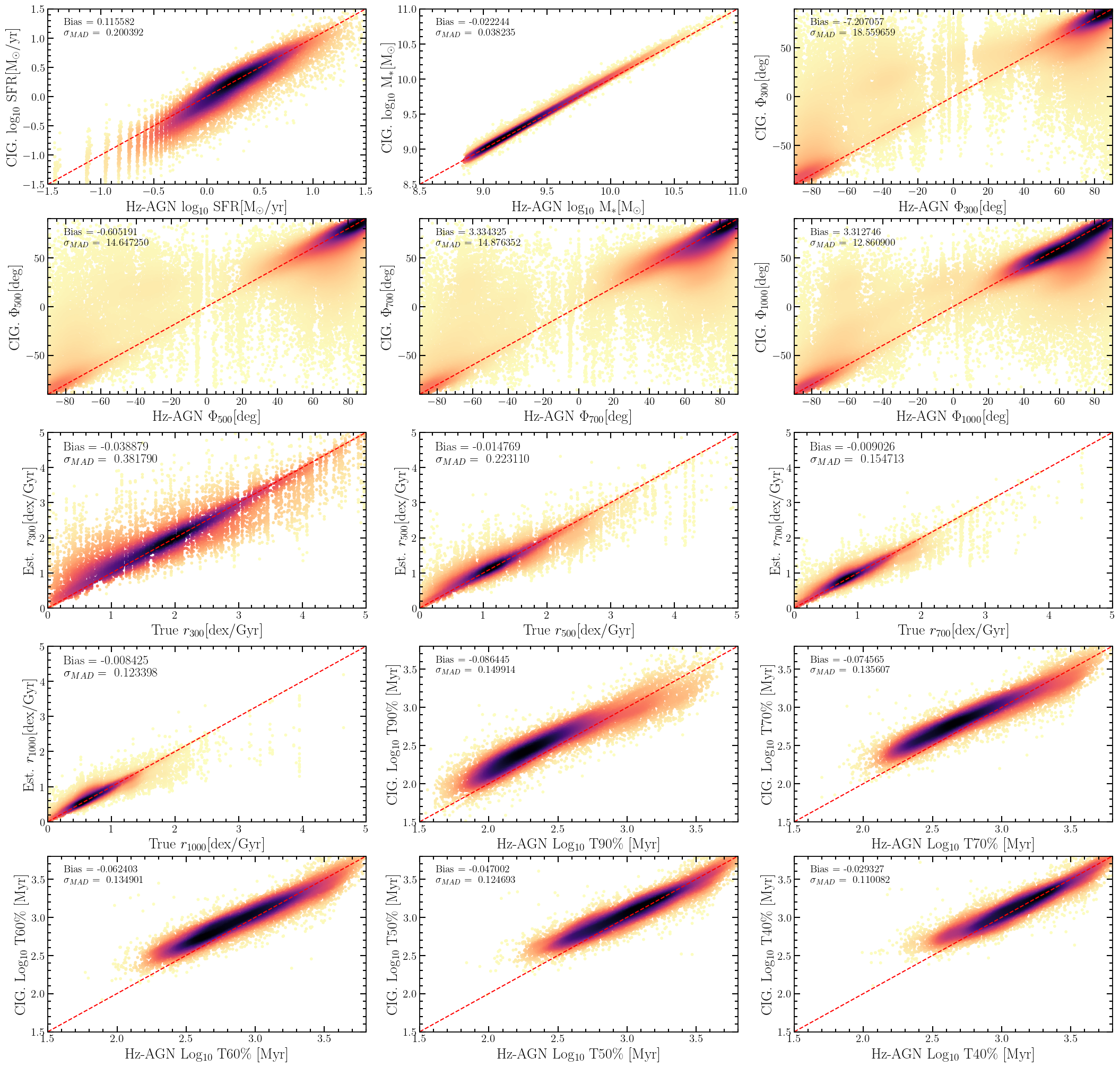}
       \caption{Results of the comparison between the intrinsic physical parameters from the {\sc Horizon-AGN} simulation and the \cigale{} SED modeling using the non-parametric SFH model. The reference values established from the simulation are shown on the $x$-axis, and the results from fitting the {\sc Horizon-AGN} catalogs are shown on the $y$-axis. The description of the different panels is the same as in Fig.\ref{mock_N-levels}.}
       \label{HZ-CIG}
\end{figure*}

To rigorously assess the accuracy and reliability of the parameters measured by \cigale{}, we include an additional test based on a hydrodynamic simulation. The {\sc Horizon-AGN} (Hz-AGN) simulation~\citep{dubois14, laigle19} is a state-of-the-art cosmological simulation that incorporates detailed models of galaxy formation and evolution, generating a large diversity of complex SFHs making it a robust framework for comparing derived parameters against true values. The advantage of this simulation is that the template library included in \cigale{} is established independently from Hz-AGN, and doesn't share the same complexity in terms of SFH, metallicity, or dust attenuation.

\subsection{The Horizon-AGN mock catalog}
The Hz-AGN simulation is based on the adaptive mesh refinement code \texttt{RAMSES} \citep{teyssier02} and incorporates a comprehensive suite of physical processes including gas cooling down to a temperature 10$^{4}$\,K following from H and He collisions including metals contributions \citep{sutherland-dopita93}, gas heating from a uniform UV background \citep{haardt-madau96} for a reionization redshift at $z=10$. Star formation is modeled following a Kennicutt-Schmidt law \citep{schmidt1959, kennicutt1998} with a constant star formation efficiency of 0.02. Stellar feedback is implemented via the mechanical energy \citep{dubois08} coming from stellar winds and supernovae type II from the \texttt{STARBURST99} model \citep{starburst99}, as well as the contribution of supernova type I explosion rate following \citet{greggio83}. The feedback process from AGNs is implemented on the Hz-AGN simulation as modeled in \citet{dubois12}, counting for the radio and quasar AGN modes, depending on the Eddington-limited Bondi-Hoyle-Lyttleton gas accretion rate onto massive black holes. This step is necessary to properly represent the AGN-driven early-type galaxies and the morphological diversity on cosmological simulations \citep{dubois13, dubois16}.

It has been shown that the Hz-AGN simulation reproduces reasonably well basic statistics on galaxy population (mass and \ion{HI} mass function, cosmic SFR density and morphology) \citep[see][]{kaviraj17,Kokorev2021,Margalef-Bentabol2020}, providing a wealth of information on various galaxy properties, such as stellar mass and SFR, formation ages, and redshift for galaxies between $0<z<4$. Simple stellar populations (SSP) from the \citet{BC03} model, the IMF as well as the metallicity of stellar particles in the Hz-AGN have been calibrated to match the observational signatures of the COSMOS2015 catalog \citep{laigle16}. For each galaxy, the dust column density (in front of each stellar particle which is part of it) is computed using the gas metallicity as a proxy for the dust distribution, and assuming a constant dust-to-metal mass ratio and a Milky Way dust extinction curve. Dust attenuation is therefore dependent on the geometry and metal content of the galaxy, while nebular lines were not included in the flux contamination over mock photometry. More details are provided in \citet{laigle19}. In Hz-AGN, the true SFR and stellar mass are available at every age of each simulated galaxy. It is worth noting that these `true' SFHs have a time resolution of 1\,Myr. Moreover, each stellar particle in the simulation represents $2\times10^6\,\Msol$, which sets the mass resolution of the SFH.

We used the mock catalog created by \citet{laigle19}. The integrated galaxy spectra are computed by adding all the dust-attenuated SSPs of each galaxy stellar particle. The flux is obtained by integrating the redshifted galaxy spectra through the filter transmission curves. In addition to the COSMOS filters already produced by \citet{laigle19}, the COSMOS-Web NIRCam filters were added to the mock catalog. Noise is added to the predicted flux according to the behavior of the flux uncertainties as a function of flux in the real data. In addition to the true redshift from virtual sources on the Hz-AGN simulation, photometric redshift was computed using the code \lephare{} \citep{Arnouts02,Ilbert06}. We note that we used the photometric redshifts already computed in \citet{laigle19}, without available NIRCam bands at that time. 

\subsection{\cigale{} derived parameters versus Horizon-AGN simulated data }

We systematically compare the SED-fitting results from \cigale{} with the true parameters extracted from the Hz-AGN simulation. We focus on quantities describing the star-formation activity (stellar masses, SFRs, formation epoch, and $\migvec$ components). We run \cigale{} applying the same configuration as described in Sect.~\ref{sec:cigale} and Table~\ref{cig_tab}. We set the Hz-AGN redshifts to the photometric redshifts computed with \lephare{}. The nebular emission module is disabled in this \cigale\ run to be consistent with the Hz-AGN mock photometry which does not include them. 

As described above, the simulation encompasses many galaxy populations spanning different mass, SFR, and stellar metallicity ranges, ensuring the validation covered a representative galaxy sample. This comparison tests the impact of photometric noise on the physical parameter recovery, but also the effect of complex SFHs, continuous metallicity enrichment, and inhomogeneity in the dust distribution. Since we adopt the same IMF and the same BC03 SSP within \cigale{} and Hz-AGN, we do not test the impact of such a choice in the modeling. The method to extract the photometry, observational limitations (e.g., PSF variation), as well as the impact of nebular emission is not tested. A modified \citet{calzetti2000} attenuation law is implemented as well, with a UV bump at 2175\,\AA. 

Figure~\ref{HZ-CIG} shows the comparison between the parameters derived with \cigale{} and the true parameters extracted from the Hz-AGN simulation. We adopt the same parameters and definitions as in Sect.~\ref{sec:mock}. The physical parameters describing the current state of a galaxy, such as the SFR and the stellar mass, show an excellent agreement between the estimated and true values. Specifically, the stellar masses recovered from the SED-fitting show a precision better than 0.04\,dex. Then, we compare $\Mstar$ quantities related to the SFH, such as the lookback time at which a galaxy formed a given percentage of its final stellar mass, at the considered redshift. This time depends on the recovery of the SFH shape and age. In the two bottom rows of Fig.~\ref{HZ-CIG}, we compare the true and estimated times T{$i$}\%[Myr] (in log scale), with $i$ ranging from 40\% to 90\%. We obtain a remarkable agreement between the true and the SED-fitting derived values for the formation epoch, considering a cumulative stellar mass higher than 40\% of the total stellar mass. The bias on the lookback time (in log) at which a galaxy has assembled between 40\% to 90\% of its stellar mass is 0.03 and 0.07\,dex respectively, and $\sigma_{\textrm{MAD}}$ varying between 0.11 and 0.15\,dex. We stress that for all percentiles, at values of log$_{10}$T${i\%}\leq$2.5 ($\approx 300$ Myr), there is a non-negligible bias of approximately 0.1 dex. This bias can be attributed to the effects of positive-only inference, where the inferred parameter is constrained to be non-negative. As a result, the posterior probability distribution is skewed, often with a long tail extending toward higher values. The mean of the distribution is systematically biased due to the asymmetry in the posterior, particularly when the true value is near the lower bound. This reflects the sensitivity of the mean to the distribution’s shape, leading to an overestimation of the inferred parameter. Nevertheless, to be conservative, we limited our analysis to timescales greater than 300 Myr to mitigate the impact of this bias.

As the main analysis of this work relies on the measurement of the galaxy migration within the $\SFRM$ plane, we also quantified the precision on the recovery of the $\migvec$ components, $\Phidt$ and $\rdt$ (see the definition in Sect.~\ref{non-param}). We computed this parameter over the last 300, 500, 700, and 1000\,Myr using the true Hz-AGN SFHs and compared it with the values retrieved by \cigale\ on reconstructed SFHs. We note that the choice of the timescales at which the migration vector is computed is based on the time intervals between a set of consecutive redshift bins (see Sect.~\ref{0-4} and Fig. \ref{ms_prog}, \ref{pass_prog}, \ref{fig_summarize}).  

Our comparison reveals that the migration vector parameters $r_{700}$ and $\phi_{700}$ are well-recovered across the full range of stellar mass, SFR, and redshift. However, the recovery of $\sigma_{\phi_{700}}$ becomes more challenging at higher redshifts ($z > 1.5$) and for galaxies with lower SFRs, particularly those around the MS. This is consistent with the increased uncertainties in SFR and stellar mass estimates in these regimes, which are inherently more difficult to constrain. These findings highlight the robustness of our method in recovering the main trends in galaxy migration, while also identifying specific regimes where the reconstruction of $\sigma_{\phi_{700}}$ is less reliable.

We also find that the bias on the migration angle depends on the angle itself, as already shown and discussed in Sect. \ref{sec:mock}. Consequently, the $\Phidt$ values indicative of a declining SFH (rising) could be even more declining (rising) in reality. We note that the distribution of $\Phidt$ presents a bimodality, as the one found in the mock catalog based on COSMOS-Web data (see Sect.~\ref{sec:mock}). We interpret this bimodality as the combination of galaxies that are quenching with rapidly declining SFH, and star-forming galaxies with a significant $\Phidt>30^\circ$ which is necessary to maintain them on the MS (see Sect.\ref{disc:MS}).

The tests presented in this section already integrate the impact of uncertainties associated with photometric redshifts (photo-z). We compared migration vectors derived from true redshifts and photometric redshifts to measure how they impact the reconstruction of the SFHs. Our findings revealed that while photo-z estimates introduced some scatter in the migration vectors, the overall trends remained consistent with those measured based on true redshifts. We checked that 10\% of catastrophic failures in the faintest magnitude bins is sufficiently small to preserve the main conclusions in our paper.

Finally, we stress that the Hz-AGN catalog may not fully replicate the complexity of real galaxies. First, the Hz-AGN catalog is limited to stellar masses above $10^{9}\, \Msol$. Secondly, we find notable differences when we compare the $\SFRM$ plane obtained using the intrinsic physical parameters from the simulation and the plane obtained with the COSMOS-Web observations. This comparison reveals notable differences in galaxy distributions, especially the lack of passive sources in the Hz-AGN simulation at high redshifts ($z>2$) (See Sec. \ref{app:hz-CWeb}, Fig. \ref{app:fig}). While Horizon-AGN serves as a powerful tool for validating our method and assessing observational uncertainties, the discrepancies with observed populations indicate that the simulation's treatment of galaxy evolution physics, particularly in regimes such as high-redshift quenching, may not fully replicate the complexity of real galaxies. This highlights the complementary role of observations in refining galaxy formation models while reaffirming the importance of cosmological simulations as a robust validation framework. These differences emphasize the importance of the migration vector formalism, as a galaxy's position in the $\SFRM$ plane alone does not fully constrain its evolutionary history in the last billion years. We note that our conclusions are similar if we use the physical parameters recovered with CIGALE rather than the intrinsic values to establish the Hz-AGN $\SFRM$ plane.

\begin{figure*}
    \includegraphics[width=1\textwidth]{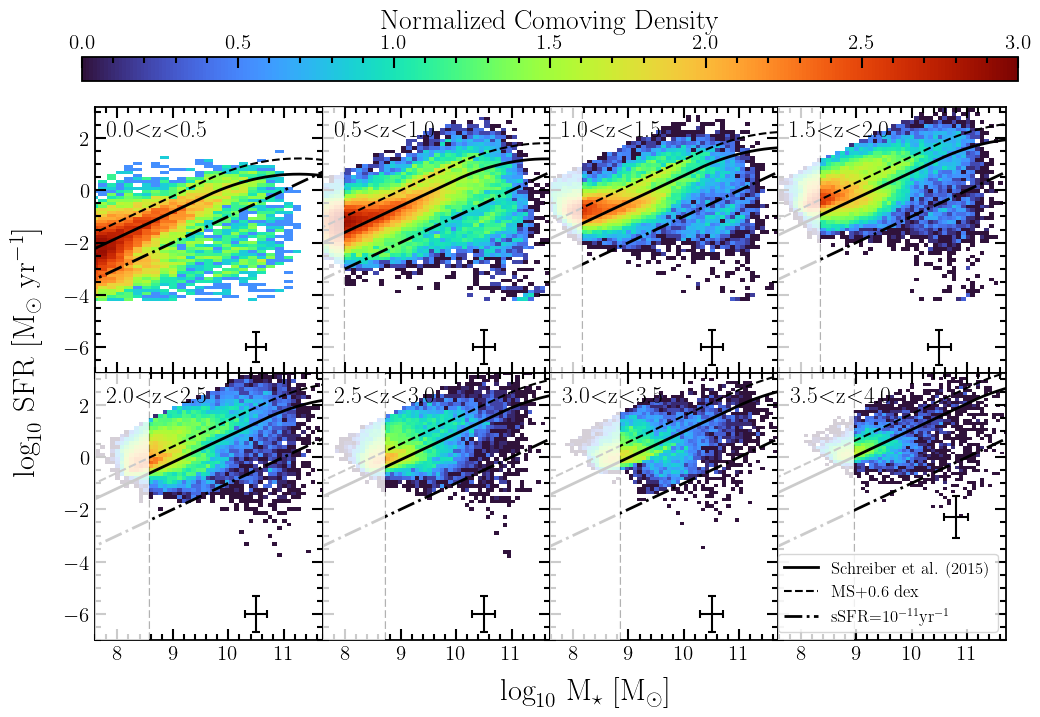}
       \caption{Two-dimensional histogram illustrating the $\SFRM$ plane across redshifts $z=0$ to 4 color coded by the density of sources over the plane. Each cell in the histogram represents the mean normalized comoving density within a particular region of the $\SFRM$ plane. Cells are displayed only when they contain at least ten galaxies, ensuring statistical robustness. Among these parametrizations, the model proposed by \cite{Schreiber15} (black solid line) presents the closest agreement with the observed dense regions within the histogram. The black dashed line presents the limit on which the SB region is considered ($\Delta {\rm MS}>0.6$\,dex, \citealp{Rodighiero2011}), while the dashed-doted black line shows the limit for the passive region ($\rm sSFR<10^{-11}\,yr^{-1}$).  The white-shaded regions indicate the mass completeness limits obtained following the method outlined by \citet{pozzetti10}.  }
       \label{ms_den}
\end{figure*}

\section{Galaxy sample\label{sec:sample}}

\subsection{Selection}

We selected a galaxy sample from the COSMOS-Web survey, spanning the redshift range from 0 to 4. We choose this redshift range to include a significant portion of cosmic history which encompasses the peak in the cosmic SFR density. We limit ourselves to $z<4$ to reach a mass completeness limit sufficient to define the MS and be able to produce a robust statistical analysis.

We implemented a selection criterion based on the reduced $\chi^2$ obtained from SED modeling with \cigale\  to include only galaxies whose observed properties closely align with our theoretical expectations, ensuring well-constrained physical parameters. We used a percentile-based selection at the 98th quantile, retaining galaxies with $\chi^2_{\text{red}}$ values below this threshold ($\chi^2_{\text{red,lim}}\approx50$). This approach mitigates potential sources of systematic error and ensures that our analysis is based on robust and reliable data. To ensure robust measurement of the flux in the rest-frame optical with a signal-to-noise ratio (S/N) greater than five, a magnitude cut is applied in the F444W filter. This corresponds to approximately 95\% completeness (see \citet{shuntov25}), resulting in a limiting magnitude of $m_{\rm F444W}^{\rm lim}=27$. The S/N > 5 threshold was chosen to ensure that the detected flux is statistically significant, reducing the impact of noise on the measurements. This magnitude cut also helps eliminate spatial selection biases caused by small variations in depth across the COSMOS-Web survey area.

\subsection{Mass completeness limits}
We adopt the method outlined by \citet{pozzetti10} to determine mass completeness limits for our galaxy sample across distinct redshift intervals. For each galaxy, we compute the stellar mass limit as the mass that a galaxy would have if its apparent magnitude was the limiting magnitude of the survey in the NIRCam F444W band, i.e., $m_{\rm F444W}^{\rm lim}=27$. We then consider the 20\% faintest galaxies and calculate the 85$^{\rm th}$ and 95$^{\rm th}$ percentile of the stellar mass limit distributions. By adopting these completeness limits, we minimize biases that could arise from losing specific populations at low stellar masses. The resulting mass completeness limits across different redshift intervals are shown in Table~\ref{mass-lims}.

\begin{table}[H]
    \centering
    \renewcommand{\arraystretch}{1.5} 
    \caption{Stellar mass completeness limits.}
    \begin{tabular}{c|c||c}
        \multirow{2}{*}{Redshift} & \multicolumn{2}{c}{log$_{10}\Mstar^{\rm lim}[\textrm{M}_\odot$]} \\
        \cline{2-3}
        & 95\% Completeness & 85\% Completeness \\
        \hline
        \hdashline
        $0 < z < 0.5$   & 7.59 & 7.44 \\
        \hline
        $0.5 < z < 1$   & 7.98 & 7.86 \\
        \hline
        $1 < z < 1.5$   & 8.15 & 8.05 \\
        \hline
        $1.5 < z < 2$   & 8.35 & 8.22 \\
        \hline
        $2 < z < 2.5$   & 8.57 & 8.41 \\
        \hline
        $2.5 < z < 3$   & 8.72 & 8.58 \\
        \hline
        $3 < z < 3.5$   & 8.86 & 8.74 \\
        \hline
        $3.5 < z < 4$   & 8.95 & 8.76 \\
    \end{tabular}
    \label{mass-lims}
\end{table}

\subsection{Impact of photometric redshift uncertainties}

Uncertainties in the photo-z estimate are a critical factor in the SFH reconstruction. The comparison between spectroscopic and photometric redshifts in COSMOS-Web shows an excellent agreement between both estimates, with less than 10\% of failures for the faint sample $m_{{\rm F444W}}>25\,{\rm mag}$. This result is consistent with the known photo-z uncertainties in the COSMOS-Web dataset (Shuntov et al., in prep), even lower because we limit our comparison to the mass-complete sample. The failure rate is similar to the one seen in the simulation, with the same kind of patterns (true high-z put at lower z). To further investigate the impact of photo-z uncertainties on our results, we recomputed the SFHs after having set the redshifts to the spec-z values. While the agreement between $\Phidt$ and $r_{\Delta t}$ measured with both photo-z and spec-z is remarkable, the scatter in the inferred migration vector angles ($\sigma(\Phidt)$) is reduced by a factor of four when using spec-z. This reduction in scatter highlights the impact of photo-z uncertainties on the inferred angles. We still find a population of outliers. Less than 10\% of sources have $\Phidt$ which differs by more than $\pm\, 30^\circ$ with the reference angle measured using spec-z. These galaxies are distributed along the MS and do not create a spurious population of SB or passive galaxies region where we found a small impact from photo-z uncertainties, as the fraction of $\Phidt$ outliers is statistically insignificant for this population. This is likely due to the uniform colors of passive galaxies, which are easier to characterize with photometric redshifts, thus minimizing discrepancies. 

While these catastrophic failures could contribute to the reported dispersion in the migration vector angle in the MS, their effect is not expected to be statistically significant since they represent less than 7\% of the MS galaxies. 
The significant reduction in scatter and outliers when using spec-z suggests that photo-z uncertainties are indeed an important source of uncertainty in this analysis, particularly for the inferred angles and likely for the magnitudes as well.
Despite uncertainties in the photo-z regarding the migration vector, we showed with the Hz-AGN simulation that we reached the precision necessary to extract the overall trends presented in this paper. Additionally, when considering a z-spec data set, even with significantly reduced uncertainties, the results may still be non-representative due to the lack of diverse sources, such as faint passive galaxies \citep{khostovan2025}.

\section{Results\label{sec:results}} 

In this section we first describe the distribution of the galaxies of the COSMOS-Web sample in the $\SFRM$ plane. Then, we study the migration of these galaxies within this plane based on their SFHs, investigating possible coherent patterns in their evolution. 

\subsection{The COSMOS-Web $\SFRM$ plane}\label{MS}

In Fig.~\ref{ms_den}, we show how galaxies are distributed over the $\SFRM$ plane at different redshift bins. 
The MS is seen up to $z=4$, our earliest redshift bin. We test four different MS models from the recent works of \citet{speagle14, Schreiber15, leja22}, and \citet{popesso23}. The \citet{Schreiber15} parametrization is the most consistent with the observed MS (MS$_{\textrm{obs}}$) across different redshift ranges (the only one shown for clarity). To quantify the difference between the \citet{Schreiber15} parametrization (MS$_{\textrm{Sch}}$) and the MS$_{\textrm{obs}}$, we select star-forming galaxies over the rest-frame (NUV-r) vs. (r-Ks) color-color diagram as described in \citet{moutard16}. We fit the distribution of SFRs at different stellar mass and redshift bins using a log-normal distribution. We analyze the deviation by looking at the $\rm MS_{\textrm{Sch}}-MS_{\textrm{obs}}$ difference. For galaxies at $0<z<2$, we find no significant deviation, with $\vert \rm MS_{\textrm{Sch}}-\rm MS_{\textrm{obs}}\vert <0.10$\,dex. At high redshift $2<z<4$, the difference can reach $\rm \vert M_{\textrm{Sch}}-\rm MS_{\textrm{obs}}\vert\lesssim 0.15$\,dex for the most massive galaxies. Given this analysis, the relatively low differences between MS$_{\textrm{Sch}}$ and our observed MS, and a non-negligible bias (>0.2 dex at z>2 at the high-mass end) from the others aforementioned parametrizations, we set the model from \citet{Schreiber15} as a reference to define different regions on the $\SFRM$ plane.

\begin{figure}
    \includegraphics[width=1\columnwidth]{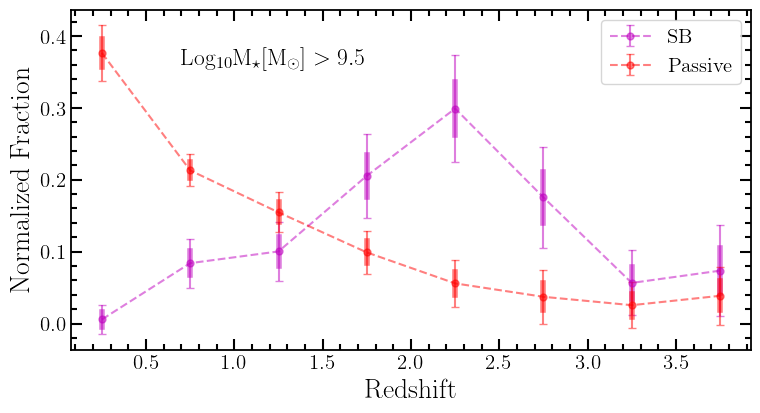}
       \caption{Fraction of galaxies categorized as SB and passive as a function of redshift. The fractions are computed with respect to the entire (z$<4$) dataset of galaxies (in a given redshift bin), with stellar masses $>$10$^{9.5}\,\Msol$. SB are galaxies that lie at more than $0.6\, \rm{dex}$ above the MS \citep{Rodighiero2011} and passive galaxies have an ${\rm sSFR}<10^{-11}\,\rm yr^{-1}$. Candle sticks represent the statistical errors in the fractions, while the overlaid error bars account for systematic uncertainties due to the adopted MS definition (obtained by changing our MS reference by $\pm$0.2 dex). Colors are magenta for SB and red for passive galaxies.}
       \label{frac_ms}

\end{figure}

We now consider the SB population, defining them as galaxies with an SFR which is 0.6\,dex (a factor 5 in SFR) or more above the MS relation at a given stellar mass and redshift, an arbitrary limit as in \citet{Rodighiero2011}. As seen in Fig.~\ref{ms_den}, the occupation of the SB region of the $\SFRM$ plane varies with redshift. To quantify this we show in Fig.~\ref{frac_ms} the relative fraction (to the total number of galaxies) of galaxies more massive than ${\rm M_\star}>10^{9.5}\,\Msol$ in the SB region as a function of redshift. Using our definition, the fraction of SB galaxies is less than 10$\%$ at $0<z<1.5$. At $1.5<z<2.5$, the SB fraction increases up to $30\%$, and eventually decreases at $z>3$ down to approximately $10\%$.

We now consider galaxies with an sSFR (defined as the ratio between the SFR and the $\Mstar$)  below $10^{-11}\rm \,yr^{-1}$ as passive \citep[e.g.,][]{ilbert10}. 
In Fig.~\ref{ms_den}, we see how this population increases with cosmic time, which is particularly visible at $\Mstar\gtrsim10^{9}\,\Msol$. As shown in Fig.~\ref{frac_ms}, from redshifts 0 to 4, the fraction of passive galaxies rapidly increases with cosmic time, reaching more than $35\%$ at $0<z<0.5$. At higher redshifts, this fraction decreases to $11\%$ at $1.5<z<2$, down to $5\%$ at $2.5<z<3$. At redshifts $z>3$ a significant portion of the galaxies (3\%) can already be considered passive based on the sSFR criteria. This result indicates a gradual quenching of star formation activity in these galaxies as they evolve, starting already above $z>4$. We will quantify the increase of this population by measuring the evolution of the galaxy's stellar mass function by type in Shuntov et al. (in prep). In the next section, we study how these galaxies migrate in this part of the $\SFRM$ plane based on their reconstructed SFH.

\begin{figure*}
\includegraphics[width=1\textwidth]{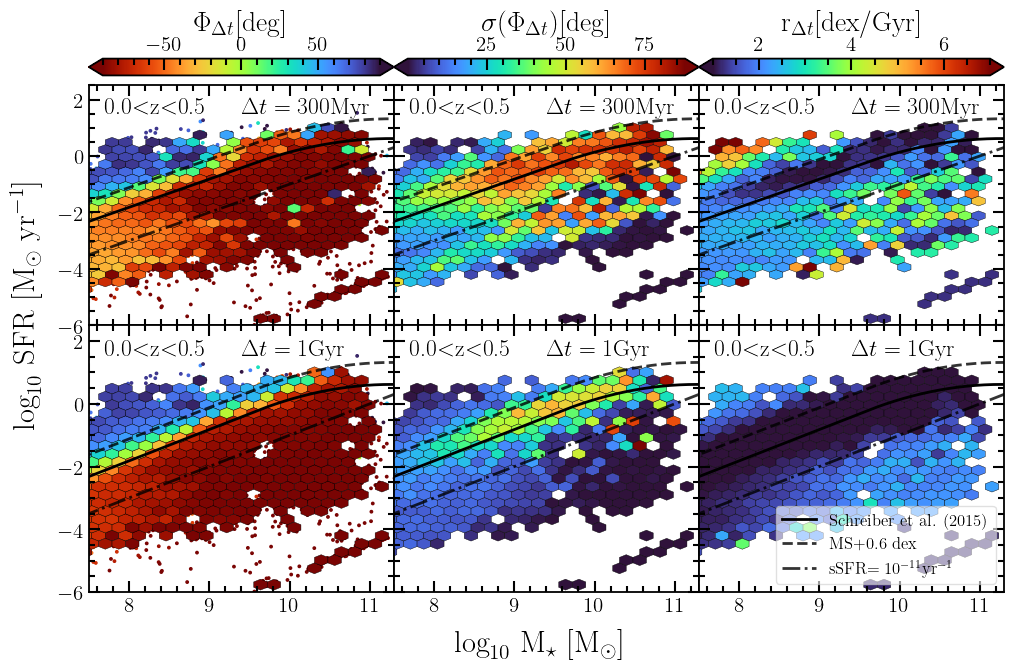}
       \caption{Three-panel visualization of the $\SFRM$ plane at $0<z<0.5$ binned in hexagonal 2D cells and color coded according to different parameters. In the left panels, the cells are color coded according to the median angle of migration ($\Phidt$) providing insights into the trajectory directions within the plane. In the middle panels, the color scheme represents the dispersion on $\Phidt$ indicating the variety of directions. The right panel, the color coding depicts the amplitude of displacement ($\rdt$) observed over the recent history of the galaxy sample. The upper and lower panels correspond to a chosen timescale $\Delta t$ of 300\,Myr and 1000\,Myr, respectively.}
       \label{ms_vect}
\end{figure*}

\subsection{Detailed analysis of the SFHs at $0 < z < 0.5$} \label{res:0-0.5}

As introduced in Sect.~\ref{non-param}, the migration velocity vector $\migvec$ describes the past trajectory of a galaxy in the $\SFRM$ plane. By examining its mean value depending on the position in the plane, we expect to gain insight into the processes that have led galaxies to their current position with respect to the MS and quantify when galaxies transition from and into the SB and passive regions.

For the sake of clarity, we first focus on a specific redshift bin ($0<z<0.5$) to present our results. This analysis is extended to the full redshift range in the next section. We analyze the median angle $\Phidt$, its scatter, and the velocity norm $\rdt$ for several timescales between 300 and 1000\,Myr. Figure~\ref{ms_vect} shows the $\SFRM$ plane binned in hexagonal 2D cells (with the number of sources by cell >10) and color coded according to the three parameters, with $\Delta t$ = 300 and 1000\,Myr, for the upper and lower panels respectively. We selected these two timescale values as the most extreme among the ones we investigated in this analysis.
 \begin{figure}
    \includegraphics[width=1\columnwidth]{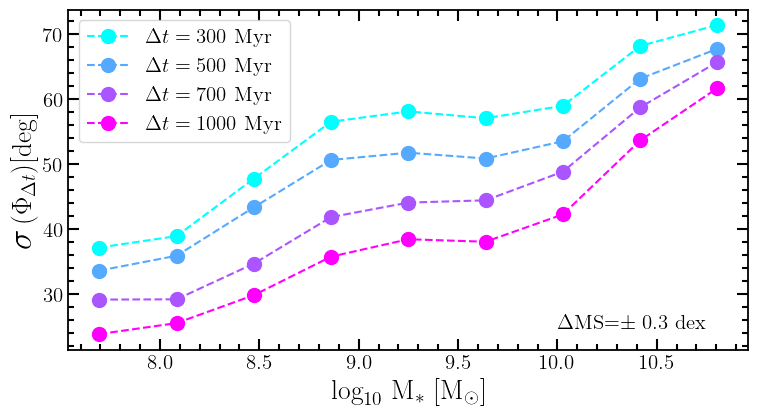}
       \caption{Dispersion of the migration angle as a function of stellar mass. This representation illustrates the evolution of the dispersion with lookback time, as a function of stellar mass for galaxies inside the MS. }
       \label{frac_ms_mass}
\end{figure}
\begin{figure}
    \includegraphics[width=1\columnwidth]{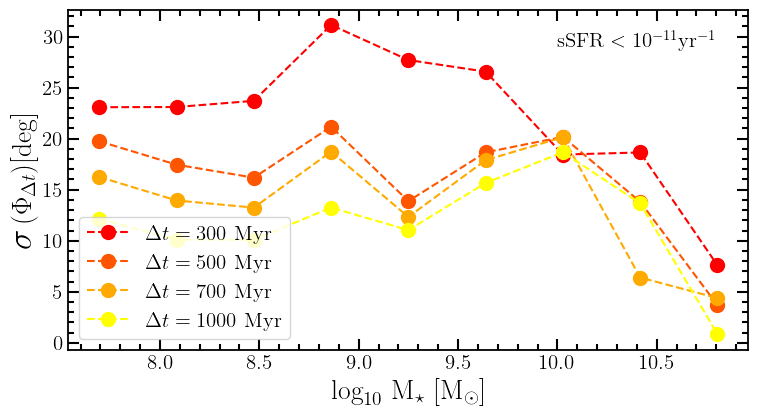}
       \caption{Dispersion of the migration angle as a function of stellar mass in the passive galaxies region.}
       \label{frac_ms_pass}
\end{figure} 
We first investigate the degree of coherence in the recent SFH in a given position of the $\SFRM$ plane. The cells in the left panels in Fig. \ref{ms_vect} are color coded according to the $\Phidt$ median values. We obtain a remarkable correlation between the positions of galaxies in the $\SFRM$ plane and their migration angles. This representation of the plane shows a well-defined stratified distribution of $\Phidt$, with a continuous decrease of its median value from the SB region toward the passive one. The cells within the same $\Phidt$ range are well aligned with the slope of the MS. For instance, cells with a value between $-50^\circ<\Phidt<-20^\circ$ overlap with the star-forming MS established in Sect.~\ref{MS}. 
The cells in the middle panels of Fig.~\ref{ms_vect} are color coded according to the dispersion of the migration angle, $\sigma (\Phidt)$. It quantifies the degree of coherence in the orientation of $\migvec$. In this representation, we find a low dispersion value for SB and passive regions, while the dispersion increases significantly toward the high mass end of the MS. We stress that one needs to be careful in analyzing the dispersion. By definition, extreme values on $\Phidt$ must be associated with lower $\sigma(\Phidt)$ as by construction $-90^\circ\leqslant\Phidt\leqslant90^\circ$. In the right panels, the cells are color coded with the median value of the velocity norm, $\rdt$, i.e., the norm of $\migvec$ normalized by $\Delta t$ (Eq. \ref{grad_sfr} and \ref{norma}). We also find a high degree of homogeneity when considering the cells parallel to the MS. Galaxies positioned along the MS demonstrate markedly low levels of velocity norm $\rdt~(\lesssim 4$ and $2$\,dex/Gyr at 300 and 1000 Myr on $\Delta$t, respectively). 
Therefore, the first conclusion of our work indicates a high degree of homogeneity of the median migration vector, with its properties depending on the distance perpendicular to the MS. Therefore, the bulk of the MS will remain unchanged as it evolves around this redshift, but its normalization will not.

Secondly, we investigate the SB region as defined in Sect.~\ref{MS}. This region presents highly positive $\phigiga$ median values ($\gtrsim$ 60°), low dispersion ($\lesssim$ 20°) and high values of $\rgiga$ ($>3$\,dex/Gyr).
\begin{figure*}
    \includegraphics[width=1\textwidth]{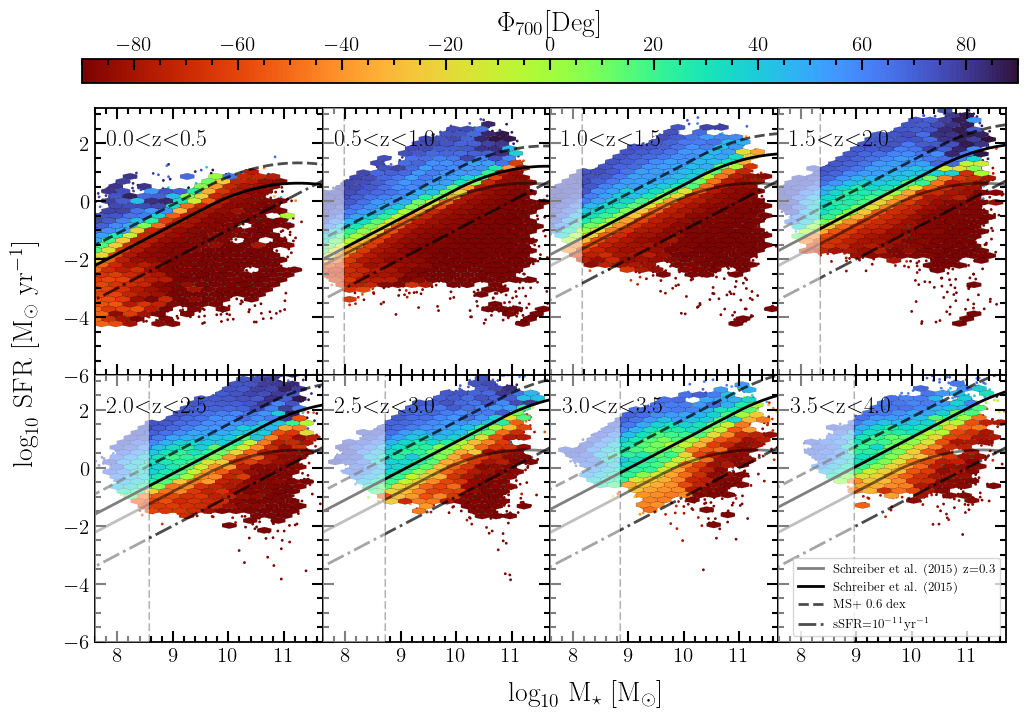}
       \caption{Plane of $\SFRM$ in different redshift bins from $z=0$ to $z=4$. The color code indicates the median value on $\Phidt$ with $\Delta t=700$\,Myr and binned in hexagonal 2D cells. The black solid line presents the MS from the model of \cite{Schreiber15} at each redshift. In each panel, the gray solid line shows the MS model at $z=0.25$. The dot-dashed and dashed lines indicate the limit to select passive and SB galaxies.} 
       \label{ms_angle}
\end{figure*}

These values are characteristic of strongly rising SFHs, as shown in Fig.~\ref{grad_sfh}. This high homogeneity of the SFHs indicates that most of the galaxies do not stay in this SB state since the fraction of galaxies with $\Phidt<0^\circ$ remains below 2\% in this region. The direction and high-velocity norm of $\migvec$ suggest that these galaxies may have even originated from the MS at higher redshifts and have moved toward the SB region at the current redshift, as we discuss and quantify in Sect.~\ref{sec:disc}.
Third, we study galaxies within the MS, within $\pm 0.3$\,dex from the MS of \cite{Schreiber15}. The dispersion of $\Phidt$ inside this region suggests that these galaxies exhibit more SFHs diversity than galaxies outside the MS. However, there is a low-velocity norm of the migration vector which implies a coherent evolution of the bulk of star-forming galaxies. In other words, galaxies from the MS may have originated from different positions within the $\SFRM$ plane, but they did not move with a velocity higher than $\rgiga \sim 1$ dex/Gyr. We interpret this result as oscillations of the galaxies within the MS. It results in the bulk of galaxies moving coherently, preserving the MS. 

Figures~\ref{ms_vect} and~\ref{frac_ms_mass} show that the general tendency on the dispersion ($\sigma(\Phidt)$), increases as $\Delta t$ decreases in the MS. This is expected due to the star-formation stochasticity probed by lower timescales. Additionally, regardless of the timescale, the dispersion within the MS increases with stellar mass. Therefore, the diversity of SFH increases significantly with a stellar mass within the MS -given the considered timescale, we do not probe short-term star-formation stochasticity but long-term (>300Myrs) trend in the SFH.\\

\begin{figure*}
    \includegraphics[width=1\textwidth]{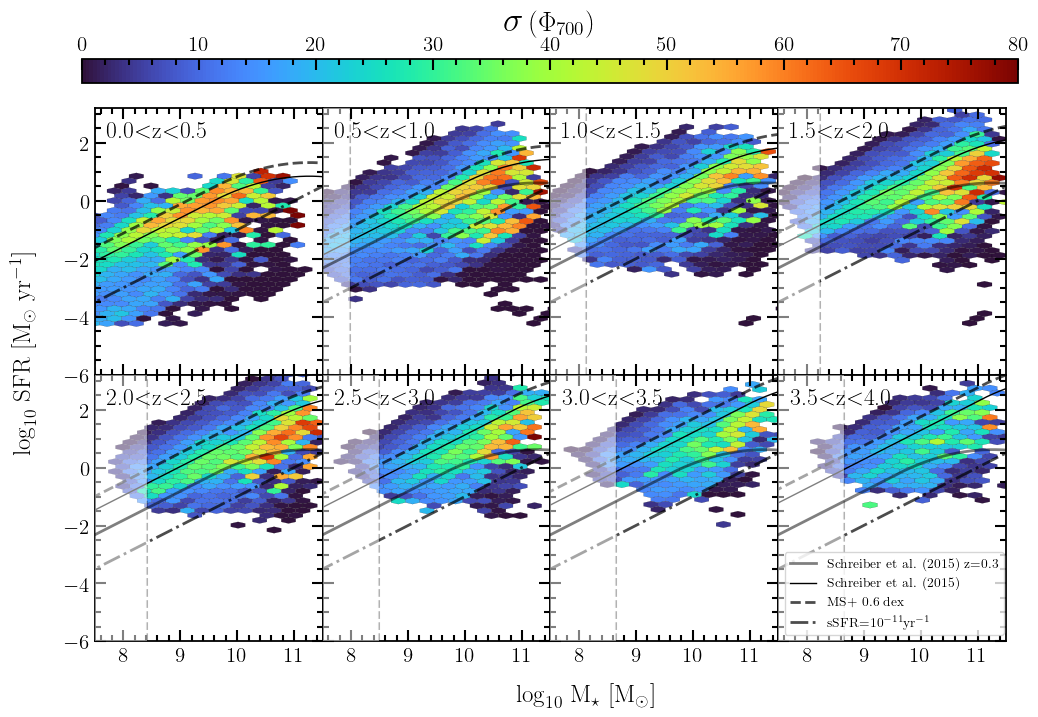}
       \caption{Same as Fig.~\ref{ms_angle} except that the color code indicates the median value of the dispersion on $\Phidt$.}
       \label{ms_disp}
\end{figure*}

\begin{figure*}
    \includegraphics[width=1\textwidth]{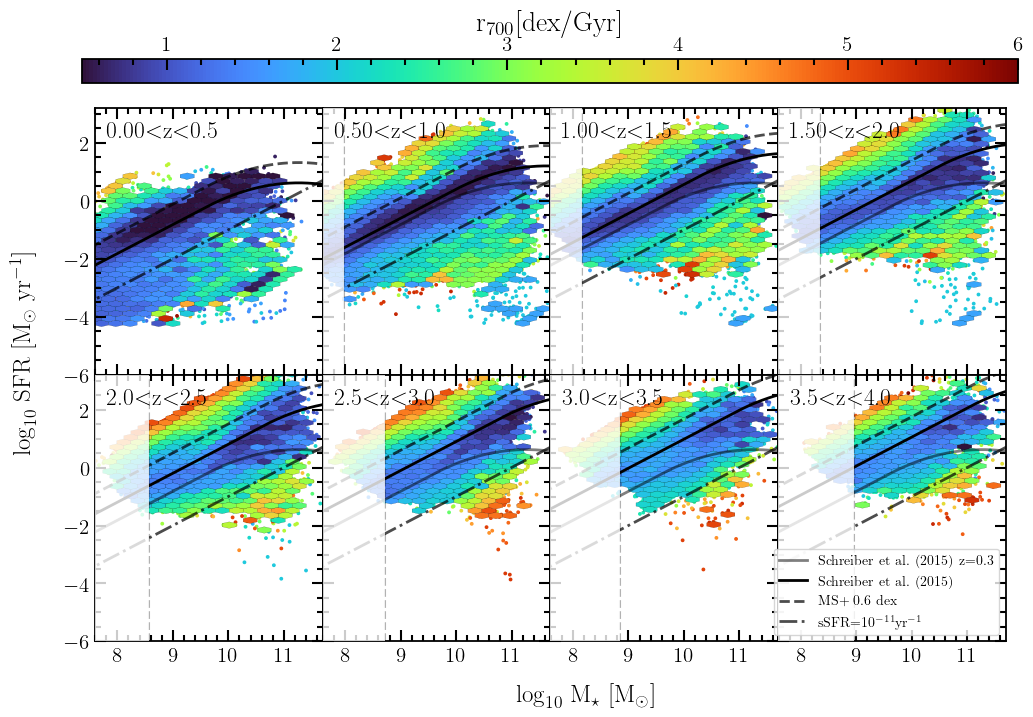}
       \caption{Same as Fig.~\ref{ms_angle} except that the color code indicates the median value of $\rdt$.}
       \label{ms_displ}
\end{figure*}
 We find that galaxies located in the passive region have migration angles with negative values, indicating a decrease in their star formation within the considered timescales, i.e., $\Delta t$ from 300 to 1000\,Myr. Massive galaxies ($\Mstar>$10$^{10}\,\Msol$), show angles $\Phidt$ < $-50^\circ$, with associated dispersion below $20^\circ$ and velocity norms going from 2 to 4 dex/Gyr, this meaning that this massive sample is dominated by galaxies that have rapidly quenched in the last Gyr, by having moved by more than 2\,dex in the $\SFRM$ plane. We discuss the origin of this population in Sect.~\ref{disc:quench}. Finally, we notice that passive galaxies with stellar masses between 10$^{7.5}$ to 10$^{9}\,\Msol$ present a larger $\sigma(\Phidt)$ than the most massive galaxies, showing more diversity in their SFH. We quantify this trend in Fig.~\ref{frac_ms_pass}, with a clear drop in $\sigma(\Phidt)$ above $\Mstar> 10^{10}\,\Msol$. 

Our conclusions do not vary significantly when considering different timescales $\Delta t$, from 300 to 1000\,Myr. We find a hint of lower $\sigma(\Phidt)$ when increasing the timescale, which suggests that the position of galaxies over the $\SFRM$ plane is driven by long-term variations in the SFH. 
\begin{figure*}
    \includegraphics[width=1\textwidth]{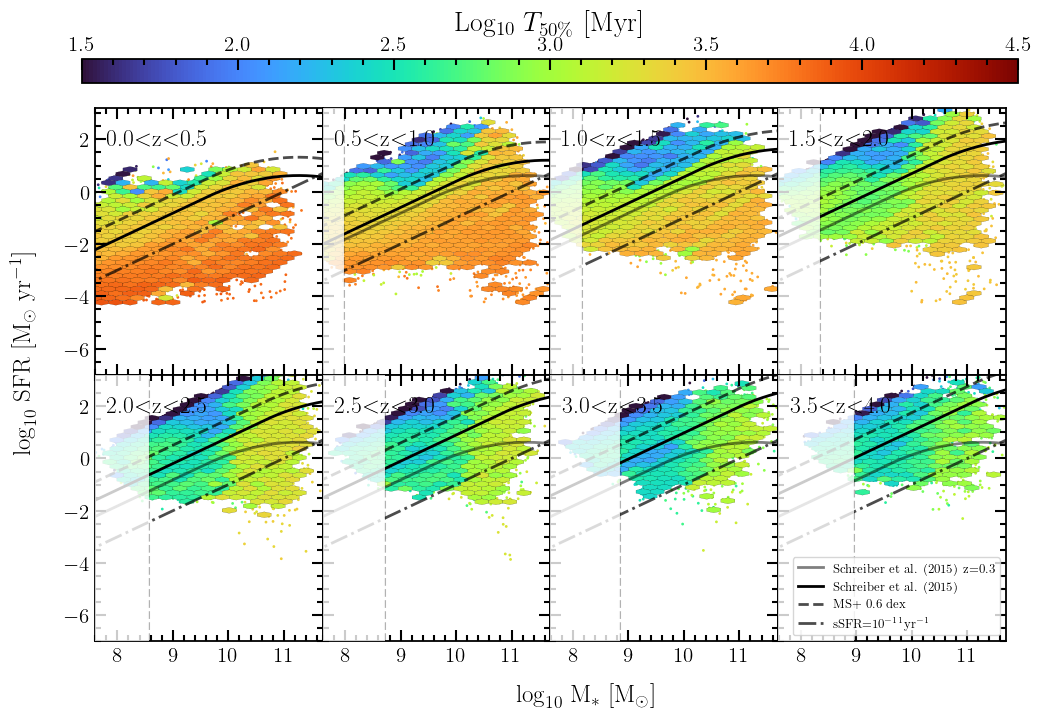}
       \caption{
       Same as Fig.~\ref{ms_angle} except that the color code indicates the median value of T$_{50\%}$.
       }
       \label{ms_t-form}
\end{figure*}

\subsection{SFHs across the redshift range $0<z<4$ \label{0-4}}

The results presented so far were focused on the specific redshift range $0<z<0.5$, testing several lookback time periods from 300 to 1000\,Myr. Since the results do not depend significantly on the considered timescale (as illustrated in Sect.~\ref{res:0-0.5}), we set $\Delta t=700$\,Myr for $\migvec$ and extend our analysis over the full redshift range $0<z<4$. Figures~\ref{ms_angle}, \ref{ms_disp}, and \ref{ms_displ} shows the median value of the migration angle $\Phidt$, its dispersion, and the velocity norm, $r_{\Delta t}$ respectively, in various redshift bins. Several trends observed at low redshift ($0<z<0.5$) can be generalized across the full redshift range ($0<z<4$).
We can generalize the coherent patterns on $\phiseven$ seen at $0<z<0.5$ to the full redshift range. In every redshift bin, we find well-stratified values of the median $\phiseven$ with an orientation parallel to the MS slope. At any redshift, the median value of $\phiseven$ decreases continuously from the SB region to the passive one. The dispersion of the migration angles is the lowest within the most extreme regions (SB and passive galaxies), while the norm is the largest. In the following, we analyze in detail each region of the $\SFRM$ plane starting from the highest sSFR.

\begin{figure}[h]
\includegraphics[width=\columnwidth]{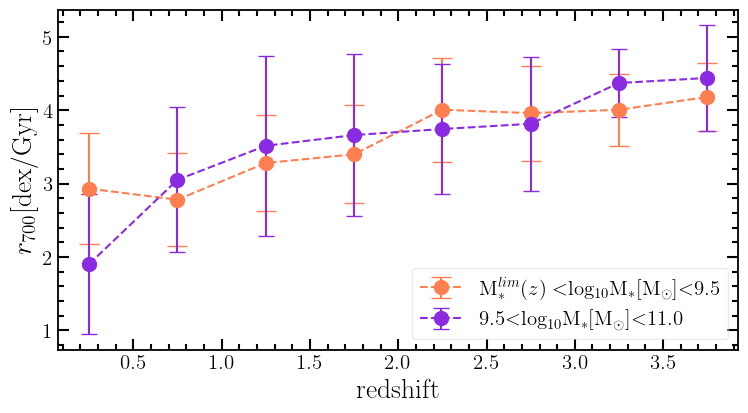}
       \caption{Median velocity norm at 700\,Myr ($\rseven$) as a function of redshift for SB galaxies. Orange connected circles are galaxies with stellar masses between log$_{10}\rm M_{\star}^{\rm lim}(z)/\Msol<\log_{10}M_\star/\Msol<9.5$, while the violet connected circles show galaxies with stellar masses between $9.5<\log_{10}\rm M_\star/\Msol<11.0$. Error bars indicate the dispersion on $\rseven$.}
       \label{speed_SB}
\end{figure}

~\newline

Starburst galaxies: We first obtained that galaxies in the SB region have strongly rising SFHs with $\phiseven\gtrsim60^\circ$ and low dispersion on these measurements (<10$^\circ$). This population is dominated by galaxies having formed 50\% of their stars recently (in the last 300\,Myr, see Fig.~\ref{ms_t-form}), indicating that most of their stellar mass has been formed in a recent burst. They also present high-velocity norms over the $\SFRM$ plane with $\rseven\,\gtrsim2$ dex/Gyr. Such conclusions remain valid across all redshift bins. However, as noted in Sect.~\ref{MS}, the fraction of massive galaxies in the SB region increases with redshift only until $z=2.2$. In this population, the value of $\rseven$ is even larger than the values found at lower redshifts. Figure~\ref{speed_SB} illustrates this result, showing that the most massive galaxies reach a displacement up to $\rseven=4.20\pm0.50$ dex/Gyr at $z>3$, while low redshift $z<0.5$ massive galaxies show low displacement around $\rseven=2.90\pm0.70$ dex/Gyr.

\begin{figure}[h]
\includegraphics[width=1\columnwidth]{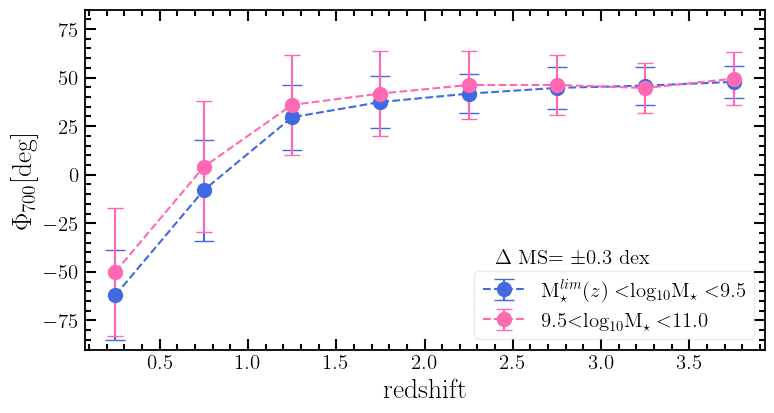}
       \caption{Median migration angle as a function of redshift for MS$\pm0.3$\,dex galaxies. Pink-connected circles indicate galaxies with stellar masses between M$_{\star}^{lim}(z)<$log$_{10}$M$_\star<9.5$, while blue-connected dots indicate galaxies with stellar masses between $9.5<$log$_{10}$M$_\star<11.0$. Error bars show the dispersion on the migration angle.}
        \label{phi_disp_ms}
\end{figure}

~\newline

Main sequence galaxies: We analyzed the galaxies encompassed within the MS region ($\pm 0.3$\,dex around the MS of \citealp{Schreiber15}). We observe a consistent trend as the one found at $0<z<0.5$: MS galaxies exhibit low to moderate median $\rseven$ values (Fig. \ref{ms_displ}). This indicates that galaxies within the MS experience limited movements in the $\SFRM$ plane as cosmic time progresses. This is true over the full considered redshift range $0<z<4$. This finding explains why this densely populated region undergoes minimal evolution over the $\SFRM$ plane. Such stability is essential for maintaining the structural integrity of the MS, as discussed in Sect.\ref{disc:MS}. Another interesting trend of the MS galaxies is the continuous change of the median value of $\phiseven$ along cosmic time. Measurements of the angle $\phiseven$ inside the MS show a shift from declining SFHs at low redshifts to rising ones at high redshifts (see Fig.~\ref{ms_angle}). Figure~\ref{phi_disp_ms} summarizes the evolution with cosmic time of $\phiseven$ for both, low mass (log$_{10}$M$_{\star,\rm lim}(z)/\Msol<$log$_{10}\Mstar/\Msol<9.5$) and more massive MS galaxies ($9.5<$log$_{10}\Mstar/\Msol<11.5$), with error bars illustrating the dispersion on the migration angle, $\textrm{\large $\sigma$}\left(\phiseven\right)$. At redshift between $0<z<1$, the MS region is populated by galaxies showing a migration angle within the range $-70^\circ<\phiseven<4^\circ$, characteristic of galaxies presenting a declining recent star-formation activity. As we rewind in cosmic time, the distribution of angles inside the MS shifts continuously toward higher $\phiseven$ values. At redshift between $2<z<4$, the angle values reach a plateau and are relatively constant ($30^\circ<\phiseven<70^\circ$) corresponding to a rising SFH.

~\newline

\begin{figure} \includegraphics[width=1\columnwidth]{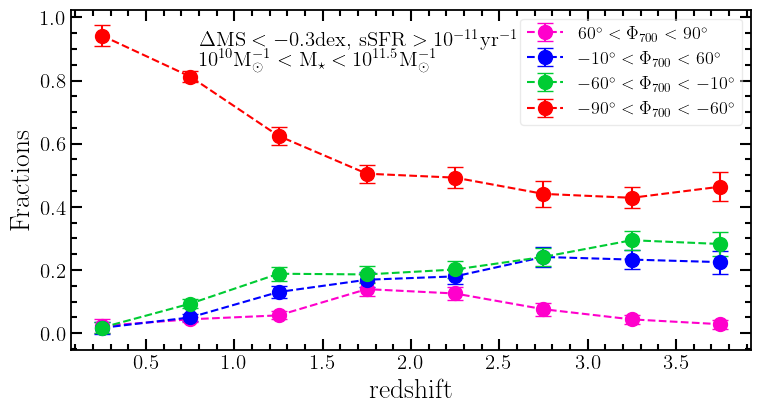}
       \caption{Fraction of galaxies as a function of redshift relative to the region over the $\SFRM$ plane defined by $ \Delta MS <-0.3 \rm{dex}$, $\rm{sSFR}^{-1}>10^{-11}\rm{yr}$, and $10^{10}< \Mstar[\Msol] < 10^{11.5}$. Fractions are discretized by the recent trends on their SFHs, showing recently rising SFHs ($60^\circ<\phiseven<90^\circ$), recently flat SFHs ($-10^\circ<\phiseven<60^\circ$), slowly declining SFHs ($-60^\circ<\phiseven<-10^\circ$), and fast declining SFHs ($-90^\circ<\phiseven<-60^\circ$), respectively colored in magenta, blue, green, and red. Uncertainties on the fraction were obtained as in Tab. \ref{Tab:fracs}}
       \label{frac_gv}
\end{figure}

Green valley galaxies: The galaxies selected in the region below the MS ($\Delta$MS$<-0.3$\,dex) but not passive ($\rm{sSFR}>10^{-11}\rm{yr}^{-1}$) are located in the region of the \(\SFRM\) plane generally called the GV  \citep{salim2007,whitaker2012,quilley22}. Although the GV is considered to be a transition region from the star-forming region to the passive one, some studies suggest that it could also be a place where we can find galaxy rejuvenation \citep{Schawinski2014,Chauke2019,Belli2019}. We observe in Fig.~\ref{ms_disp} that massive GV galaxies present the highest dispersion in $\phiseven$ at all redshifts. They are also characterized by a low-velocity norm $\rseven$. We show in Fig.~\ref{frac_gv} the fraction of GV galaxies depending on their migration angle. The GV area is dominated by galaxies with a declining SFH, consistent with starting their transition toward the passive region. Still, we find a significant population of galaxies with rising SFHs (magenta-connected circles) at intermediate redshifts, representing up to $\sim20\%$ at intermediate redshifts $1.5<z<2.5$ and in the mass range $10^{10} < \Mstar/\Msol < 10^{11.5}$, which could be associated with rejuvenation. 

~\newline

Passive galaxies: As pointed out in Sect.~\ref{MS}, there is a notable increase with cosmic time in the density of galaxies transitioning into the passive phase. We find a population at $3.5<z<4$ with strongly declining SFH ($\phiseven<-60^\circ$) and rapid migration within the $\SFRM$ plane ($\rseven>3$ dex/Gyr). Therefore, the quenching process initiates early in the age of the Universe and occurs first in the most massive galaxies with $\Mstar\gtrsim10^{10}\,\Msol$. This population grows in density with cosmic time, indicating a continuous quenching of the massive galaxies. We discuss in detail in Sect.\ref{disc:quench} what could be the progenitors of this passive population. While the quenching occurs first within the massive galaxies, we find an increase of passive galaxies with $10^{8} < \Mstar/\Msol < 10^{9}$ which appears at $z<1$. As discussed in Sect.~\ref{res:0-0.5}, this population presents similar SFH characteristics as other passive galaxies ($\phiseven<-60^\circ$, $\rseven>3$ dex/Gyr), but a higher dispersion of $\phiseven$ than their massive counterparts. 

Finally, we discuss a general trend in the galaxy ages across the $\SFRM$ plane. Figure~\ref{ms_t-form} presents the $\SFRM$ plane colored by the parameter \( T_{50\%} \), which represents the lookback time at which 50$\%$ of a galaxy's total stellar mass was formed. We find that the most massive galaxies tend to have higher \( T_{50\%} \) values than low-mass galaxies, with an abrupt change above $10^{10}\,\Msol$ and $z>1.5$. This indicates that massive galaxies formed their stellar mass earlier in the cosmic timeline. This pattern aligns with the downsizing phenomenon described by \citet{Cowie1996}, where more massive galaxies form their stars earlier and faster than less massive ones. This result is also consistent with \citet{Aufort24} who show that star-forming galaxies present clear mass-dependent SFH-based machine-learning techniques applied to COSMOS2020 \citep{weaver22b}. This phenomenon suggests that massive galaxies experience intense and frequent star formation episodes at higher redshifts followed by a drop in star formation, leading to higher \( T_{50\%} \) values. Mergers could play a role in building the mass and in the quenching process, which cannot be tested with our dataset. In contrast, low-mass galaxies generally form their stars over more extended periods, resulting in lower \( T_{50\%} \) values. This highlights the differences in SFHs between massive and less massive galaxies.

\section{Discussion}\label{sec:disc}

For two decades, the position of the galaxies in the $\SFRM$ plane has been commonly used to select galaxies with similar properties \citep[e.g.,][]{noeske2007, peng2010, popesso23}. This classification is based on the instantaneous state of the galaxies in terms of SFR and stellar mass. In this paper, we use the reconstructed SFH of individual galaxies to study their migration over the $\SFRM$ plane. Thanks to the migration vector, we can directly estimate the positions of the galaxy progenitors at earlier times and quantify how galaxies move within the $\SFRM$ plane.

\subsection{Maintaining the galaxy main sequence over cosmic time}\label{disc:MS}

As shown in Sect.~\ref{MS}, we find a well established MS out to $z=4$, well described by the \citet{Schreiber15} parametrization. The position of the MS evolves with redshift, as found in many studies  \citep[e.g.,][]{Brinchmann2004,Elbaz07,whitaker2012,popesso23}. In Sect.~\ref{res:0-0.5} and \ref{0-4}, we show that galaxies within the MS share a similar median angle $\Phidt$, meaning that MS galaxies have moved in the same direction over the $\SFRM$ plane in the last billion years. Such a condition seems necessary to explain how the slope of the MS remains constant over cosmic time \citep[e.g.,][]{speagle2014}. 

As shown in Fig.~\ref{phi_disp_ms}, the average value of $\Phidt$ within the MS has declined continuously over time. Star-forming galaxies that belong to the MS ($\pm 0.3$ dex) exhibit a rising SFH at $1.5<z<4$ which reverts to a declining SFH at $z<1.5$. Such a result is consistent with \citet{pacifi13} who found that the SFHs of star-forming galaxies selected at $z<1.4$ rise and fall in a bell-shaped manner. Our sample spans a larger time range and the galaxies observed at $z>1.5$ probe the rising part of this bell-shaped SFH. While we could be tempted to link this shape with the evolution of the comoving SFR density \citep[e.g.,]{Madau14}, such analysis would require considering the comoving number density of each star-forming population contributing to the SFR density.

When studying the MS galaxies in more detail, we find that the dispersion of $\Phidt$ increases toward massive galaxies (Fig. \ref{frac_ms_mass}). This means that galaxies could come from various directions in the $\SFRM$ plane, which could fight against the existence of the MS at high masses. \citet{ilbert15} find an increase of the MS scatter at high masses in agreement with our analysis. However, this increase of $\sigma(\Phidt)$ is compensated by the low norm of the migration vector seen in Fig.~\ref{ms_displ}. Even though massive MS galaxies move in a broader range of directions than lower-mass galaxies, they cross lower distances in the $\SFRM$ plane, which preserves the MS at high mass. Indeed, considering that the dispersion of $\Phidt$ within the MS is relatively high and can reach up to 50 degrees, the MS would not persist over cosmic time if galaxies were experiencing large movements (i.e., high values of the nom $r_{\Delta t}$). Therefore, a low norm is needed to preserve the MS. Indeed, the test described below shows that our reconstructed SFH are consistent with the observed MS evolution.

\begin{figure*} \includegraphics[width=2\columnwidth]{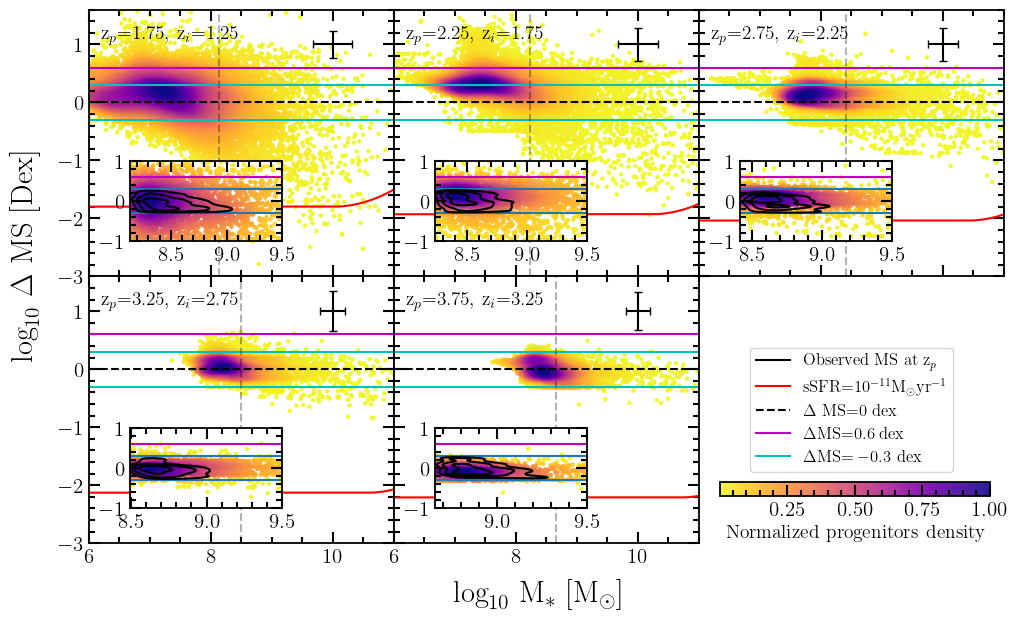}
       \caption{Distance to the MS ($\Delta \textrm{MS}$) as a function of the stellar mass (in log scales). Scatter color-coded points represents the SFR and $\Mstar$ expected for progenitors at z$_p$ = \{1.75, 2.25, 2.75, 3.25, 3.57\} obtained by computing the expected change in SFR and $\Mstar$ of MS galaxies ($\Delta$MS=$\pm0.3$ dex) at $z_i$ = \{1.25, 1.75, 2.25, 2.75, 3.25\}. Black contour lines in the inset panels show the observed (mass complete) MS galaxies at z$_p$. The color bar represents the normalized progenitor density over the plane at z$_p$. The evolution in time of galaxies over the $\SFRM$ plane is predicted via the $\migvec$ components by computing the change in SFR and $\Mstar$. The dashed lines correspond to the MS of \citet{Schreiber15} estimated at z$_p$.}
       \label{ms_prog}
\end{figure*}

To go one step further, we can use the migration vectors to understand from where MS galaxies originate. We select galaxies within $\pm 0.3$\,dex around the MS in a given redshift bin $z_1<z_{\rm i}<z_2$. Based on the migration vector of individual galaxies, we can reconstruct the expected position of their progenitors at earlier epochs corresponding to a redshift noted $z_{\rm p}$. We used the following relation $\Delta log_{10}{\rm SFR}=\rdt \times dT \times \sin(\Phidt)$ and $\Delta$ log$_{10}\Mstar=\rdt\times dT \times \cos(\Phidt)$, where $dT$ is the cosmic time elapsed between the redshift of the galaxy ($z_i$) and $z_{\rm p}$. We choose $z_{\rm p}$ to be the mean of the nearest redshift bin $z_2<z_{\rm p}<z_3$. We also choose $\Delta t$ (within the range 300 to 1000 Myr) to be the closest as possible to $dT$ to minimize extrapolation. Therefore, we can compare the expected positions of the progenitors of MS galaxies at $z_1<z_{\rm i}<z_2$ with the MS position at $z_{\rm p}$ expected from the \citet{Schreiber15} parametrization. Figure~\ref{ms_prog} illustrates this method at several contiguous redshift bins. For instance, black dots on the upper left panel of Fig.~\ref{ms_prog}, show MS galaxies selected at $1<z<1.5$ (with $z_{\rm i}=1.25$). Then, their expected positions on the $\SFRM$ plane at $z_{\rm p}=1.75$ were recovered via the individual migration vectors, showing the expected distribution of these progenitors (color-coded dots) respectively to the MS from \citet{Schreiber15}, at $z_{\rm p}=1.75$. We find that $\approx 70\%$ of MS galaxies' progenitors were already in the MS. Only $5\%$ came from the SB region and none from the passive region. The inset panel in Fig.~\ref{ms_prog} shows the observed (mass complete) galaxies at $1<z<1.5$ showing good agreement with the \citet{Schreiber15} parametrization. A significant population of the MS ($\sim$5\%) originates from the massive end of the GV. Such results corroborate the possible rejuvenation discussed in Sect.~\ref{0-4} and/or a larger diversity of SFH in massive MS galaxies. We note that the positions of the progenitors can be reconstructed well below the mass completeness limit of observed galaxies. 

We extend this analysis at all redshifts when the cosmic time elapsed between two redshift bins remains below 1\,Gyr to allow for a robust reconstruction, i.e., above $z>1$. Over the considered range of redshifts, we find that the fraction of MS galaxies that already belonged to the MS in the previous redshift bin is encompassed between 64\% and 83\%. A maximum of 5\% originates from the SB region and none from the passive region. So, we conclude that most of the MS galaxies remain within the MS as time evolves, while a small fraction of them ($5\%$) experienced an SB phase. We also conclude that the migration vector which describes the movement of galaxies in the $\SFRM$ plane, can explain coherently the observed evolution of the MS from \citet{Schreiber15}.

As an additional test, we extended the previous analysis for all galaxy populations, i.e., not limiting ourselves to the MS galaxies. We select galaxies at $z_i$ and infer their position in the $\SFRM$ plane at an earlier time ($z_p$ with $z_p>z_i$). A correlation between stellar mass and SFR is clearly visible in the reconstructed plane, which corresponds well to the observed MS at $z_p$. The offsets between the reconstructed and observed MS are lower than 0.2 dex in the mass-complete regime. The figure is not shown in this paper since the results are very similar to the one shown in Fig.~\ref{ms_prog}. Thus, we can conclude that the observed evolution of the MS matches well the expected evolution from the individual SFH, supporting the robustness of the migration vectors when used to derive average trends.

\subsection{The buildup of the passive population}\label{disc:quench}

We found 110 galaxies in the passive region of the $\SFRM$ plane at $3.5<z<4$ (Sect.~\ref{MS}). Their stellar masses are within the range $10^{10.5} <\Mstar/ \Msol<10^{11.5}$, which corresponds to the most massive galaxies at this epoch (see Fig.~\ref{ms_den}). As expected, these galaxies present a declining SFH with $\Phidt<-60^\circ$ and among the highest norm in the migration vector (see Fig.~\ref{ms_displ}). Their SFH indicates that they have formed in a burst which could be fitted by an exponentially declining function with e-folding time $\tau\approx$ 200\,Myr. According to their \T50, we find half of these passive galaxies already assembled 50\% of their stellar mass 1200\,Myr after the Big Bang (800\,Myr for three of them). Therefore, these galaxies would have formed and quenched between $5<z<7$. Recent results from several JWST surveys discovered the presence of galaxy candidates potentially as massive as $\Mstar>10^{10}\,\Msol$ at $z\gtrsim10$ \citep[e.g][]{labbe23,Casey2024}. \citet{shuntov25} find evidence for a significant population of galaxies more massive than $\Mstar>10^{10.5}\,\Msol$ at $z>5$ in the same field. Given their density, these galaxies are potentially the progenitors of our passive sample. Our sample indicates that quenching already occurs at $z\gtrsim 5$. While still extremely unclear, physical processes such as AGN feedback could be able to quench massive galaxies as early as 1.5 Gyr in cosmic time \citep{Dubois2013highz,Saxena2024,Xie2024}. 

\begin{figure*}    
\includegraphics[width=1\textwidth]{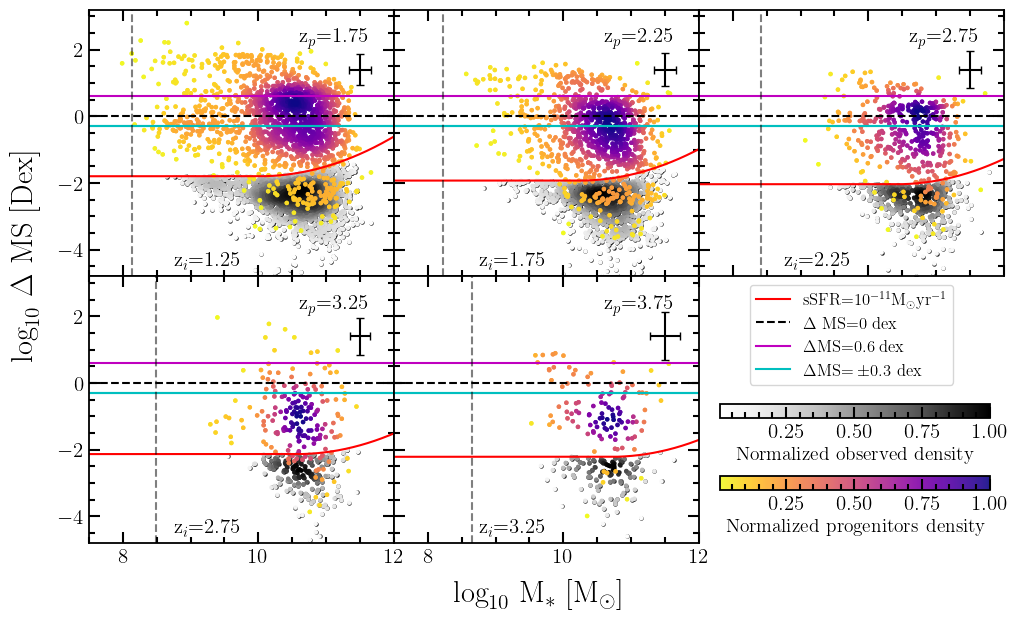}
       \caption{Distance to the MS ($\Delta \textrm{MS}$) as a function of the stellar mass (in log scales). Shaded black dots represent the observed passive galaxies at $z_i=\{1.25,1.75,2.25,2.75,3.25\}$,  while color-coded dots (by density over the plane) show the progenitors predicted MS distance at $z_p=\{1.75,2.25,2.75,3.25,3.57\}$. Progenitors' SFR and $\Mstar$ were obtained by computing the expected change in SFR and $\Mstar$ of galaxies at $z_1$. The evolution in time from $z_2$ to $z_1$ over the $\SFRM$ plane was predicted as in Fig.~\ref{ms_prog}. Cyan solid lines indicate the MS$\pm0.3$\,dex, the magenta solid line the MS+0.6\,dex, and red solid line the ${\rm sSFR}=10^{-11}\,\rm yr^{-1}$ limit for passive galaxies. The color bar represents the normalized progenitor's density over the plane at z$_p$, while the gray color bar shows the normalized observed density of sources at z$_i$}
       \label{pass_prog}
\end{figure*}

The number density of passive galaxies increases continuously from $z=4$ to $z=1$ for stellar masses above $10^{10}\,\Msol$ (Fig.~\ref{ms_den}). Such evolution is expected from the evolution of the galaxy stellar mass functions \citep[e.g.,][]{ilbert10,weaver22b, valentino23} showing a rapid increase in the number density of this population. One question is the origin of these massive galaxies which quench rapidly between $z=4$ to $z=1$. For instance, a critical question is whether passive galaxies transition directly from the MS or undergo an SB followed by rapid quenching \citep{Hopkins2008}. The rise in massive galaxies in the passive region coincides with the decline of massive SB galaxies seen in Fig.~\ref{frac_ms}, notably around $z=2$, which suggests a potential link. We follow the same method as described in Sect.~\ref{disc:MS} to study the progenitors of passive galaxies. The top-left panel of Fig.~\ref{pass_prog} shows the passive galaxies at $1<z<1.5$ (black shaded dots) and the expected position of their progenitors at $z=1.75$ relative to the MS of \citet{Schreiber15}. Most of the progenitors are massive galaxies above $\Mstar>10^{10}\,\Msol$, but they are scattered over a wide range of SFR. Around 16\% of the progenitors of the passive galaxies at $z=1.25$ come directly from the SB region at $z=1.75$, with stellar masses above $10^{10}\,\Msol$. It corresponds to the most massive SB seen in Fig.~\ref{ms_den}, which implies that part of the massive SB appears to move into the passive region. So, these SB can quench in less than 1.1\,Gyr, in agreement with a rapid quenching achieved by AGN feedback in bright mode \citep{Hopkins2008}. Still, the majority of the progenitors of passive galaxies originate from the MS (23\%) and the GV (44\%). Therefore, the main channel for the formation of these passives is probably smoother \citep[see][]{Wang2019,Williams2019,Valentino2020,Long2023,Manning2024}. Results are similar in all the considered redshift bins, while we emphasize that the time interval in the two highest redshift bins is only 300\,Myr. The GV region at $\Mstar>10^{10}\,\Msol$ contains a mix of MS galaxies, galaxies undergoing rejuvenation, and galaxies in transition due to quenching. It explains the high dispersion seen in $\Phidt$ (see Fig.~\ref{ms_disp}), with star-forming galaxies mixed with galaxies in transition to the passive region. A more detailed study of this region of the $\SFRM$ plane would be crucial to understanding physical processes leading to quenching.

The passive galaxies at $\Mstar >10^{10}\, \Msol$ are dominated by declining SFH, with a low dispersion in $\Phidt$ values (see Fig.~\ref{ms_angle} and Fig.~\ref{ms_disp}). Such homogeneity means that when galaxies start to quench, the process does not stop and the quenching continues to act. The most massive galaxies at $0<z<0.5$ with the lowest sSFR are supposed to have quenched among the earliest in the Universe, as confirmed by their \T50 ages above 10 Gyr. Even these galaxies are dominated by declining SFH.  Therefore, a physical process is needed to prevent the rejuvenation of star formation in these already passive systems. This behavior could indicate quenching processes linked to the halo mass, where quenching in massive halos is maintained by virial shocks and radio-mode AGN feedback preventing gas from being accreted onto galaxies \citep[e.g.,][]{croton2006,Bower2006, Cattaneo2006,dubois10,Beckmann2017}. Given the continuous growth of dark matter halo, this quenching mode is maintained for several Gyr which could explain the continuous decrease of star formation. 

Finally, it would seem counter-intuitive that the passive galaxies do not originate from the passive region. Similarly, the value of the norm $r_{\Delta t}$ is higher than expected in the passive region given that these galaxies are supposed to not move anymore (see Fig.~\ref{ms_displ}). We interpret this result as an indication that the passive population is dominated by galaxies that have recently fallen into this region. The comoving number density of passive galaxies has increased by more than a factor of 10 between $z=3$ and $z=1$ \citep[e.g.,][]{weaver22b}. Therefore, the contribution to the overall trend is likely outweighed by the more numerous star-forming galaxies which continuously quench and fall into the passive region. We conclude that newcomers to the passive region dominate the main trends at $z>1$. At lower redshift, newcomers are predominantly galaxies with a mass of $\Mstar< 10^{10}\Msol$, which corresponds well to the galaxies with higher $r_{\Delta T}$ values.

\begin{table*}[]
\centering
\caption{Main sequence and passive galaxy progenitors fraction over the $\SFRM$ plane.}

\label{Tab:fracs}
\begin{tabular}{|c|cccc||cccc|}
\hline
                                  & \multicolumn{4}{c||}{Progenitors of the observed MS}                                                                                                                                       & \multicolumn{4}{c|}{Progenitors of the observed Passive}                                                 \\ \hline
$z_i$ ,$z_p$, $dT$[Myr]                      & \multicolumn{1}{c|}{SB\%}                & \multicolumn{1}{c|}{SF\%}                             & \multicolumn{1}{c|}{GV\%}                             & Passive           & \multicolumn{1}{c|}{SB\%} & \multicolumn{1}{c|}{SF\%} & \multicolumn{1}{c|}{GV\%} & Passive\% \\ \hline
1.25, 1.75, 1218                   & \multicolumn{1}{c|}{12.4$\pm$3.3} & \multicolumn{1}{c|}{49.9$\pm$2.5}               & \multicolumn{1}{c|}{37.4$\pm$5.7}                & 0.3$\pm$0.1        & \multicolumn{1}{c|}{7.5$\pm$2.0}   & \multicolumn{1}{c|}{23.0$\pm$1.1}   & \multicolumn{1}{c|}{49.4$\pm$7.6}   & 14.9$\pm$1.8        \\ \hline
1.75, 2.25, 790                       & \multicolumn{1}{c|}{18.5$\pm$4.8}  & \multicolumn{1}{c|}{53.1$\pm$2.8}                & \multicolumn{1}{c|}{28.0$\pm$4.6} & 0.5$\pm$0.2    & \multicolumn{1}{c|}{6.7$\pm$2.1}   & \multicolumn{1}{c|}{20.1$\pm$1.1}   & \multicolumn{1}{c|}{51.0$\pm$7.9}   & 17.2$\pm$2.1        \\ \hline
2.25, 2.75, 545                       & \multicolumn{1}{c|}{13.0$\pm$3.4} & \multicolumn{1}{c|}{52.8$\pm$2.7}                & \multicolumn{1}{c|}{33.8$\pm$5.4}                & 0.3$\pm$0.1        & \multicolumn{1}{c|}{4.5$\pm$1.2}   & \multicolumn{1}{c|}{22.8$\pm$1.5}   & \multicolumn{1}{c|}{52.0$\pm$8}   & 13.4$\pm$1.8        \\ \hline
2.75, 3.25, 393                       & \multicolumn{1}{c|}{16.8$\pm$4.3} & \multicolumn{1}{c|}{48.1$\pm$2.6} & \multicolumn{1}{c|}{34.6$\pm$3.5}                & 0.5$\pm$0.2        & \multicolumn{1}{c|}{1.2$\pm$0.4}   & \multicolumn{1}{c|}{10.3$\pm$2.3}   & \multicolumn{1}{c|}{62.4$\pm$9.4}   & 24.2$\pm$2.3        \\ \hline
\multicolumn{1}{|c|}{3.25, 3.75, 294} & \multicolumn{1}{c|}{26.6$\pm$6.7} & \multicolumn{1}{c|}{49.0$\pm$2.6}                 & \multicolumn{1}{c|}{23.9$\pm$3.8}                & 0.4$\pm$0.1        & \multicolumn{1}{c|}{4.3$\pm$1.6}   & \multicolumn{1}{c|}{18.5$\pm$3.1}   & \multicolumn{1}{c|}{60.8$\pm$9.2}   & 11.1$\pm$3.1        \\ \hline
\end{tabular}
\tablefoot{Regions of origin (at $z_p$) of galaxies identified as MS at $z_i$ (left part of the table) and passive galaxies at $z_i$ (right part) from Figs. 17 and 18. The table presents the percentage of galaxies coming from each region of the $\SFRM$ plane—SB, star forming (SF), GV, and passive—alongside their associated uncertainties. Uncertainties were computed as the quadrature sum of bootstrap errors and systematic uncertainties due to deviations of the MS definition ($\pm 0.2$ dex). Bootstrap errors were derived from 100 repetitions, where 80\% of the total data set was sampled at each iteration, with the observed SFR, $\textrm{M}_\star$, and migration vector randomly computed from a normal distribution with central values and standard deviations based on the \texttt{CIGALE} Bayesian-like analysis. Systematic uncertainties account for reclassification due to MS adopted and the typical bias with respect to the observed distribution (see Sec. \ref{MS}), including the definition of progenitor populations on the reconstructed $\SFRM$ plane. $dT$ reports the cosmic time elapsed between $z_p$ and $z_i$}
\end{table*}
\subsection{The starburst sequence}

Our analysis reveals significant insights into galaxy movements within the extreme regions of the $\SFRM$ plane. The SB population, characterized by recent bursts in their SFHs, forms a homogeneous group. These galaxies show coherent trajectories, suggesting that similar physical processes drive them. Starburst galaxies experience rapid increases in SFR, which could be triggered by interactions or mergers and/or gas-rich inflows \citep{Rodighiero2011,Elbaz2018}.  In Sect.~\ref{MS}, we find a fraction increase of this massive SB population with a peak at $z\sim 2$. Such an increase is consistent with the increase of the gas fraction \citep{Genzel2015} which is primarily constrained through MS galaxies, \citep{Tacconi20} and the merger rate \citep[][]{Luo2014,Rodriguez2019,Forrest2024,Stumbaug2024S}.

Galaxies with flat or declining SFHs are absent in the SB region, indicating that they do not stay in a high sSFR state for long. If they did, we would find some with lower velocity displacements or flat angles based on their SFH trends. Our results suggest that SB galaxies rapidly fall back to the MS or transition quickly into the passive population.

Low \T50 values indicate that most of their stars were formed recently. When using the migration vector to trace back the position of the progenitors back in time, exactly as done in Sect.~\ref{disc:MS}, we find that the vast majority originates from the MS at much lower masses. Between two redshift bins, the SB galaxies gain almost 2\,dex in stellar mass (for any bin between $1<z<1.5$ and $3.5<z<4$). Less than 1\% originate from the SB region and none from the passive. In the previous section, we show that part of the SB quench is directly within the passive region. The rise in massive galaxies in the passive region coincides with the decline of SB galaxies, notably around $z=2$, this correlation suggests a potential link.

\begin{figure*}
    \includegraphics[width=1\textwidth]{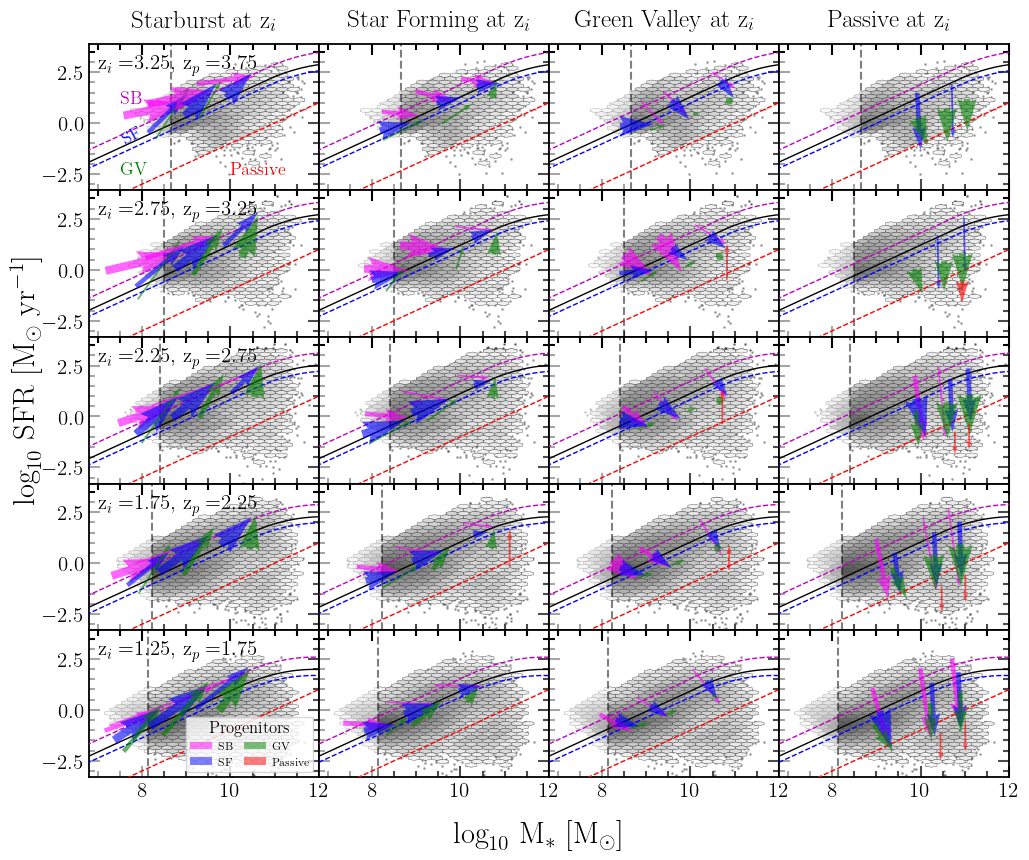}
       \caption{Schematic view of galaxy migrations across different redshift bins in the $\SFRM$ plane. The hexagonal bins represent the galaxy number density at redshift $z_i$. The arrows represent how galaxies move from a redshift z$_p$ to a redshift z$_i$. For each redshift $z_i$, we trace back the progenitors of galaxies to their positions at earlier redshifts $z_p$, using migration vector ($\migvec$) that indicates the direction and velocity of galaxy evolution over the time interval elapsed between $z_p$ and $z_i$. The arrows are color coded by the progenitor population: SB, star forming, GV, and passive. The width of the arrows is modulated by the relative fraction of sources originating from the respective regions. The head of each arrow corresponds to the galaxy positions at $z_i$, while the tail represents their positions in different regions at $z_p$. The head's arrow position in each panel is calculated for stellar mass bins equally spaced over the mass range considered at $z_i$. The solid black line shows the MS at $z_p$, with the magenta, blue, and red dashed lines indicating the boundaries for the SB, star-forming, and passive regions, respectively. Each column focuses on galaxies in different populations at $z_i$ ( SB, star forming, GV, and passive), while the rows illustrate changes for different redshift pairs $(z_i, z_p)$. The arrows highlight the quantifiable moves in the $\SFRM$.}
       \label{fig_summarize}
\end{figure*}

\section{Conclusions\label{sec:conclu}}

In this paper, we have studied the migration of the galaxies using the $\SFRM$ plane in the redshift range $0<z<4$.
We based our analysis on the COSMOS-Web survey, and our sample contained 394602 galaxies over 0.54\,deg$^2$ including flux measurements in 33 bands from the visible to mid-IR. Galaxies were selected in the NIRCam bands with $m_{\rm F444W}<27$. The associated photometric redshifts are as precise as $\sigmaz{}=0.01$ for a sample selected at $m_{\rm F444W}<24$, which degrades to $\sigmaz{}=0.025$ at $26<m_{\rm F444W}<27$. We limited our analysis to a mass-complete sample at $z<4$.

By employing sophisticated SED-fitting techniques with the code \cigale, we reconstructed the SFH of individual galaxies. Based on the SFHs, we defined a quantity named migration vector that characterizes the velocity and the direction of the galaxy movements within the $\SFRM$ for a given lookback time. We tested the robustness of the SFH reconstruction by relying on mock catalogs, especially one created with the Hz-AGN hydrodynamic simulation including complex SFHs. This comparison with simulations showed that the migration vector can be trusted for the considered lookback times between 300 and 1000\,Myr. We also found that we can reconstruct the lookback time at which galaxies assembled more than 50\% of their total stellar mass.

The galaxy distribution in the $\SFRM$ plane presents a clear MS that is consistent with the MS derived by \citet{Schreiber15}. We also found a continuous increase in the comoving density of passive galaxies with cosmic time. The SB fraction of galaxies more massive than $10^{9.5}\,\Msol$ increases with redshift and peaks at $z\sim 2$.

We used the migration vector to describe the past trajectory of galaxies in the $\SFRM$ plane. Figure~\ref{fig_summarize} summarizes our results by presenting a schematic view of how galaxies migrate from and toward different regions of the $\SFRM$ plane. We report our conclusions as follows:
\begin{itemize}
    \item Galaxies selected in the same location of the plane share very homogeneous properties in terms of recent SFH (considering a lookback time between 300\,Myr and 1\,Gyr).
    \item We find a downsizing pattern, with massive galaxies forming their stars at an earlier time and an abrupt change in the median ages above $10^{10}\,\Msol$. 
    \item Galaxies within the MS present a low migration velocity within the plane, which preserves the existence of the MS. The largest fraction of the galaxies identified as MS galaxies in a given redshift bin were already within the MS in the previous redshift bin ($>$95\% if we include the GV). Still, we find a large dispersion of $\Phidt$ that increases with stellar mass, showing that galaxies oscillate within the MS.
    \item We find a population of SB galaxies (0.6\,dex above the MS) that assembled half of their stellar mass within the last 350\,Myr. This population originates from the MS, but their stellar masses were lower by more than 2\,dex in the previous redshift bin.
    \item The GV region at $\Mstar>10^{10}\,\Msol$ presents the largest variety of SFHs, and this region probably includes a mix of MS galaxies, galaxies experiencing rejuvenation, and galaxies in transition due to quenching. 
    \item A population of passive galaxies is identified at $3.5<z<4$. Based on their SFH, they are expected to form half of their mass and quench within the redshift range $5<z<7$. 
    \item The quenching occurs first at masses above $10^{10}\, \Msol$, with a rapid increase of this population with cosmic time. We find these galaxies originate from massive regions of the $\SFRM$ plane but over a large range of SFR. A fraction as high as 15\% moves directly from the SB to the passive region in less than 1\,Gyr, showing that rapid quenching is effective. Still, the majority of passive progenitors have moved from the GV to the passive area between two redshift bins, thus showing more gradual quenching.
    \item Galaxies in the passive region of the plane show a homogeneous declining SFH over the entire considered redshift range, indicating the necessity of a physical mechanism able to quench star formation over timescales of several billion years.
\end{itemize}

This study demonstrates the potential of multi-color photometry for estimating individual galaxy SFHs. However, the quality of SFH reconstruction depends on the considered lookback time. Our tests show that we can reconstruct the bulk of evolutionary trends when considering timescales between 300 and 1000 Myr. However, our SFH reconstruction is not sensitive to shorter timescale variations, which would require additional observational diagnostics (e.g., emission lines). The upcoming compilation of spectra from over 100 surveys (Khostovan et al., in prep) and the COSMOS-3D survey's observation of over 10000 NIRCam spectra \citep[]{Kakiichi24} will significantly enrich this analysis by allowing lower timescales to be probed and burstiness to be studied. An upgraded version of \cigale{} that includes spectroscopy \citep[]{seille24} will further refine our understanding of galaxy migration along the $\SFRM$ plane.

The impact of mergers was not considered in our analysis. The SFHs derived from SED fitting provide a composite view of progenitor populations, resulting in an average migration history rather than the trajectory of individual systems. Modeling the impact of mergers in an SED-fitting approach is out of reach with current methods. Still,
mergers are modeled in the Hz-AGN simulation. Our tests show that we are still able to reconstruct the SFHs when we consider lookback times below 1 Gyr. Still, merger is a fundamental aspect of galaxy growth, and future studies will focus on integrating galaxy morphology and SFHs to enhance our understanding of how mergers shape galaxy evolution and their impact on migration patterns within the $\SFRM$ plane.
Finally, we plan to release the \cigale{} physical parameters, including the migration vectors, together with the public release of the COSMOS-Web photometric catalog (Shuntov et al., in prep). 

\begin{acknowledgements}
We would like to thank the anonymous referee for the thoughtful comments and constructive suggestions, which greatly helped to improve the clarity and depth of this manuscript. We appreciate the time and effort put into reviewing our work. We acknowledge funding from the French Agence Nationale de la Recherche for the project iMAGE (grant ANR-22-CE31-0007).
This project has received financial support from the CNRS through the MITI interdisciplinary programs and the Initiative Physique des Infinis (IPI), a research training program of the Idex SUPER at Sorbonne Universit\'e.
This work received support from the French government under the France 2030 investment plan, as part of the Initiative d’Excellence d’Aix-Marseille Université – A*MIDEX AMX-22-RE-AB-101.
We warmly acknowledge the contributions of the COSMOS collaboration of more than 100 scientists. The HST-COSMOS program was supported through NASA grant HST-GO-09822. More information on the COSMOS survey is available at \url{https://cosmos.astro.caltech.edu}. This research is also partly supported by the Centre National d'Etudes Spatiales (CNES).
This work was made possible by utilizing the CANDIDE cluster at the Institut d’Astrophysique de Paris. The cluster was funded through grants from the PNCG, CNES, DIM-ACAV, the Euclid Consortium, and the Danish National Research Foundation Cosmic Dawn Center (DNRF140). It is maintained by Stephane Rouberol. 
\end{acknowledgements}
\bibliographystyle{aa_url}
\bibliography{aanda}

\begin{appendix}

\onecolumn  
\section{$\SFRM$ plane constructed from the Hz-AGN intrinsic values and the COSMOS-Web observed photometry \label{app:hz-CWeb}} 

\begin{figure}[ht]  
    \centering
    \begin{subfigure}{0.45\textwidth}
        \centering
        \includegraphics[width=\textwidth]{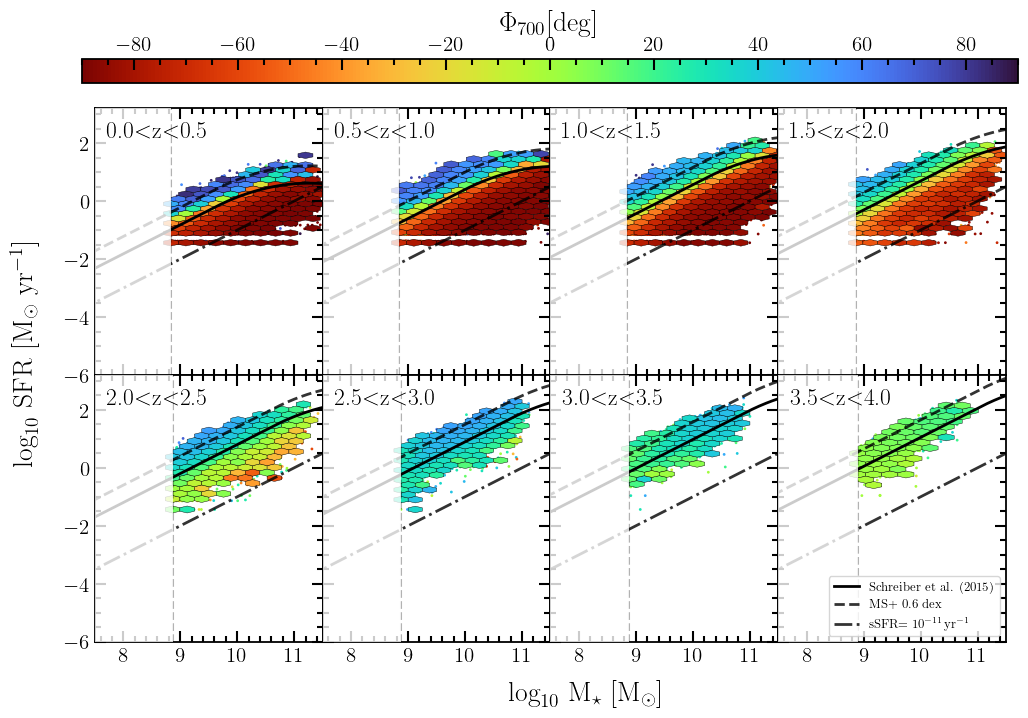}
    \end{subfigure}
    \hfill
    \begin{subfigure}{0.45\textwidth}
        \centering
        \includegraphics[width=\textwidth]{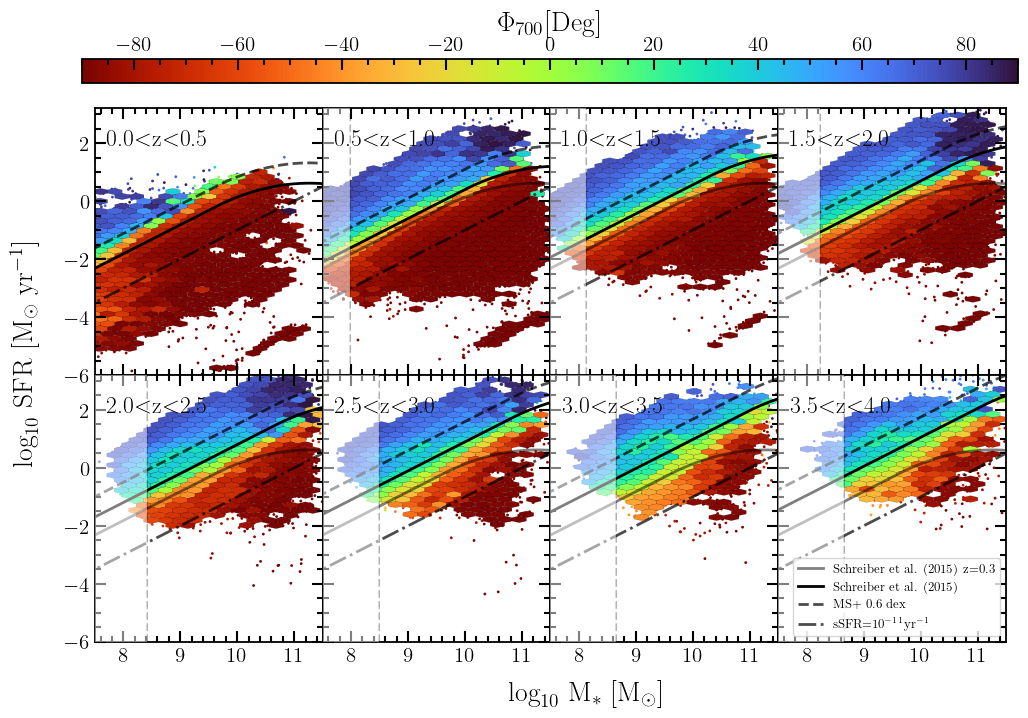}
    \end{subfigure}

    \vspace{0.3cm}

    \begin{subfigure}{0.45\textwidth}
        \centering
        \includegraphics[width=\textwidth]{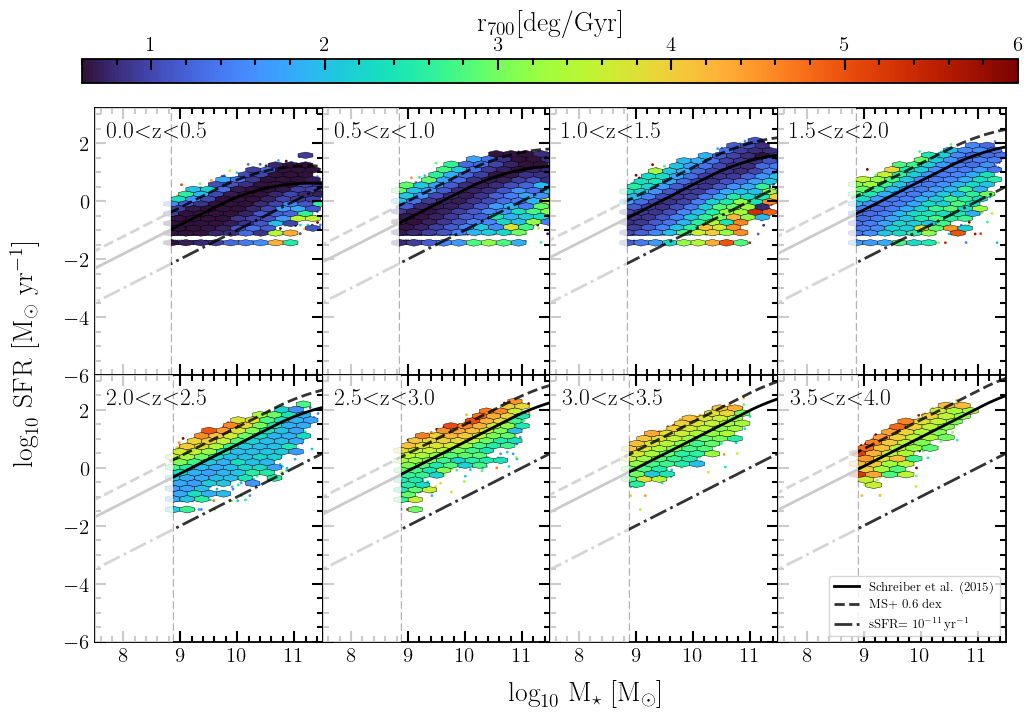}
    \end{subfigure}
    \hfill
    \begin{subfigure}{0.45\textwidth}
        \centering
        \includegraphics[width=\textwidth]{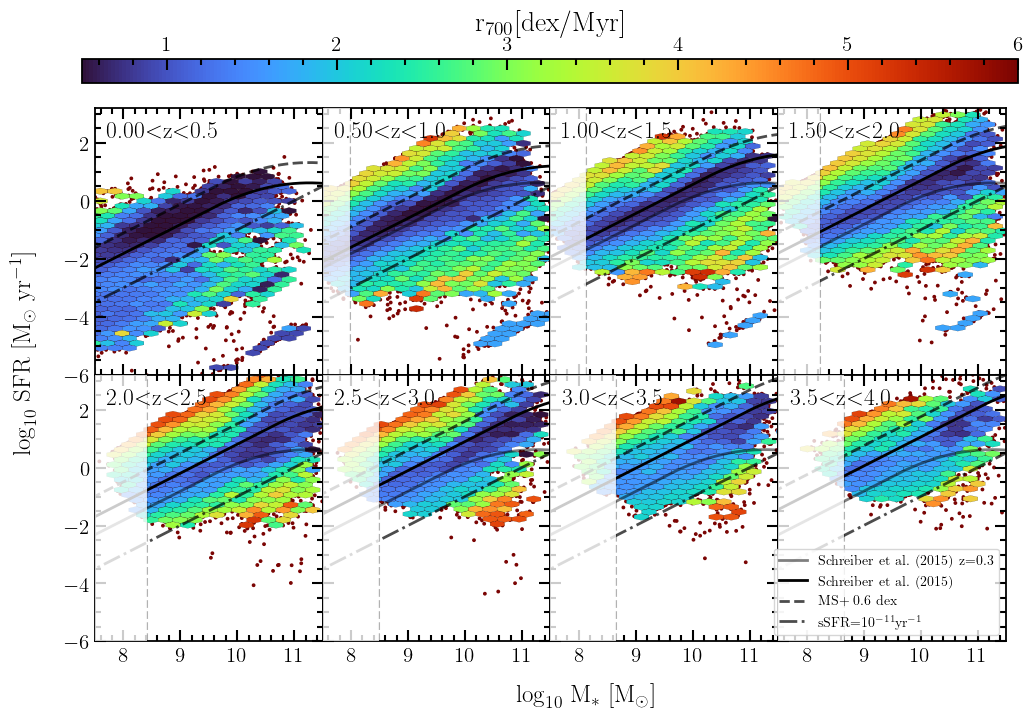}
    \end{subfigure}

    \vspace{0.3cm}

    \begin{subfigure}{0.45\textwidth}
        \centering
        \includegraphics[width=\textwidth]{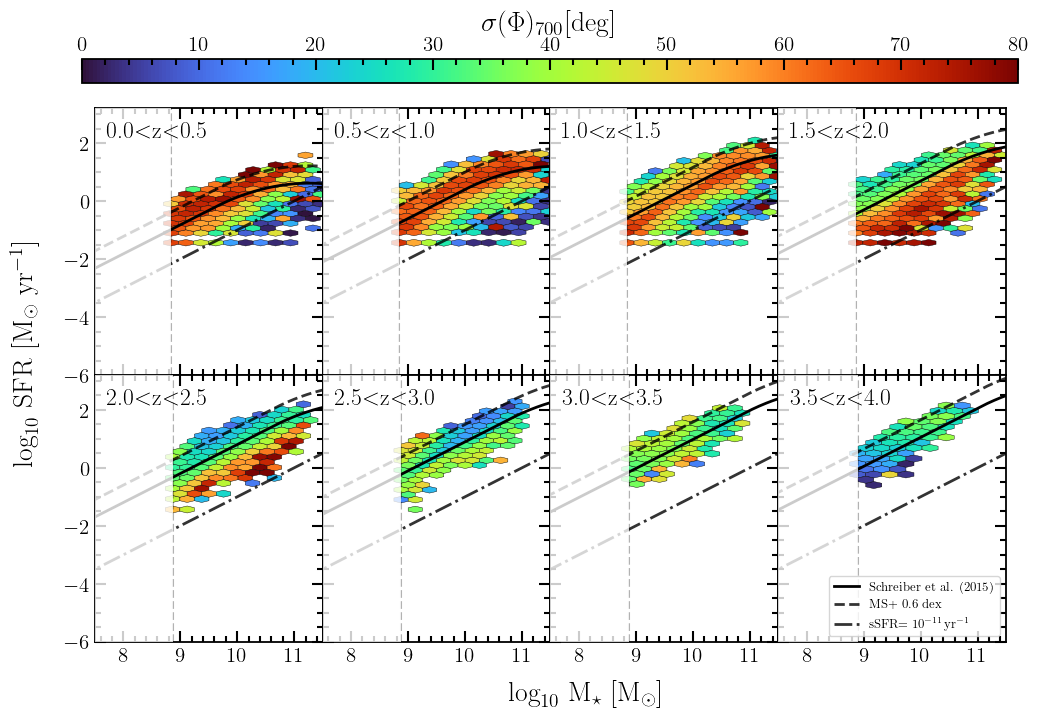}
    \end{subfigure}
    \hfill
    \begin{subfigure}{0.45\textwidth}
        \centering
        \includegraphics[width=\textwidth]{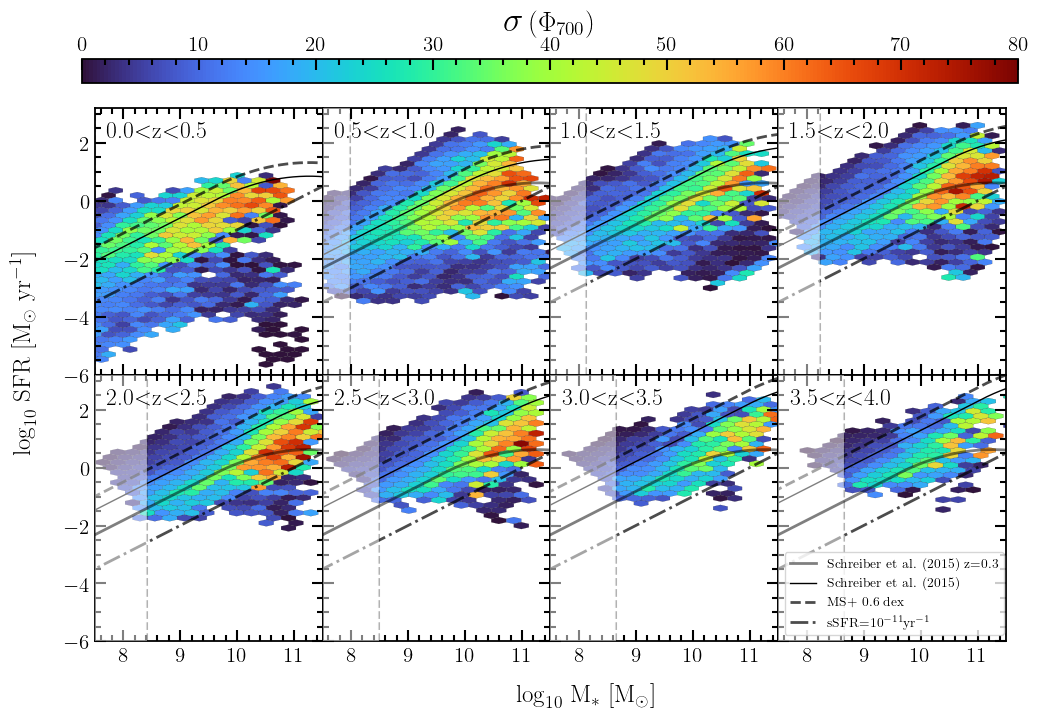}
    \end{subfigure}

    \caption{Plane of $\SFRM$ in different redshift bins ($0 < z < 4$) comparing intrinsic physical parameters from the Horizon-AGN simulation (left panels) and the COSMOS-Web observed values derived using CIGALE (right panels). 
    The top row shows the migration vector angle across the $\SFRM$ plane at different redshifts. 
    The middle row presents the migration vector magnitude, while the bottom row displays the dispersion of migration vectors.  
    The left panels are based on true SFR, stellar mass, redshift, and analytically computed migration vectors from Horizon-AGN, whereas the right panels use values derived from COSMOS-Web photometric data processed with CIGALE.  
    The color code represents the median values of migration vector components at $\Delta t = 700$\,Myr, binned in hexagonal 2D cells.  
    The black solid line represents the MS model from \cite{Schreiber15} at each redshift.  
    The gray solid line in the right panels shows the MS model at $z=0.25$.  
    The dot-dashed and dashed lines indicate the limits for selecting passive and SB galaxies.  
    White-shaded regions indicate the mass completeness limits.}
    \label{app:fig}
\end{figure}

\FloatBarrier  

\end{appendix}

\end{document}